# Non-locality and Spillover Effects of Residential Flood Damage on Community Recovery: Insights from High-resolution Flood Claim and Mobility Data

Junwei Ma[1*], Russell Blessing[2], Samuel Brody[3], Ali Mostafavi[4]

[1] Ph. D. student. Urban Resilience.AI Lab, Zachry Department of Civil and Environmental Engineering, Texas A&M University, College Station, Texas, United States.

[2] Associate Research Scientist. Department of Marine and Coastal Environmental Science, Texas A&M University at Galveston, Galveston, Texas, United States.

[3] Professor. Department of Marine and Coastal Environmental Science, Texas A&M University at Galveston, Galveston, Texas, United States.

[4] Associate Professor. Urban Resilience.AI Lab, Zachry Department of Civil and Environmental Engineering, Texas A&M University, College Station, Texas, United States.

[*]Corresponding author: Junwei Ma, E-mail: jwma@tamu.edu.

## Abstract

Examining the relationship between vulnerability of the built environment and community recovery is crucial for understanding disaster resilience. Yet, this relationship is rather neglected in the existing literature due to previous limitations in the availability of empirical datasets needed for such analysis. In this study, we combine fine-resolution flood damage claims data (composed of both insured and uninsured losses) and human mobility data (composed of millions of movement trajectories) during the 2017 Hurricane Harvey in Harris County, Texas, to specify the extent to which vulnerability of the built environment (i.e., flood property damage) affects community recovery (based on the speed of human mobility recovery) locally and regionally. We examine this relationship using a spatial lag, spatial reach, and spatial decay models to measure the extent of spillover effects of residential damage on community recovery. The findings show that: first, the severity of residential damage significantly affects the speed of community recovery. A greater extent of residential damage suppresses community recovery not only locally but also in the surrounding areas. Second, the spatial spillover effect of residential damage on community recovery speed decays with distance from the highly damaged areas. Third, spatial areas display heterogeneous spatial decay coefficients, which are associated with urban structure features such as the density of points-of-interest facilities and roads. These findings provide a novel data-driven characterization of the spatial diffusion of residential flood damage effects on community recovery and move us closer to a better understanding of complex spatial processes that shape community resilience to hazards. This study also provides valuable insights for emergency managers and public officials seeking to mitigate the non-local effects of residential damage.

# 1. Introduction

Community resilience is a complex phenomenon composed of various spatial processes and their interactions in community socio-technical systems [1-3]. One of the significant, yet less widely studied aspects of community resilience is the relationship between built-environment vulnerability and post-disaster community recovery [4]. Untangling this relationship is essential for integrating post-disaster recovery considerations during mitigation and risk reduction processes [5, 6]. Also, the characterization of local and spillover effects of built-environment vulnerability on regional recovery patterns is crucial for a better understanding of the latent spatial processes that shape community resilience to hazards [7]. This understanding is rather limited in the existing literature due to the dearth of empirical data needed to specify built-environment vulnerability and measure community recovery at a suitable scale. To address this important gap, this study harnesses fine-resolution flood damage claims and human mobility datasets in the context of the 2017 Hurricane Harvey in Harris County, Texas, to specify the extent to which vulnerability of the built environment (i.e., flood property damage) affects community recovery (based on the speed of human mobility recovery) both locally and regionally.

Using a fine-resolution flood claim dataset composed of both insured and uninsured losses, we quantify the extent of residential flood damage (as an indicator of built-environment vulnerability) across all census block groups (CBGs) in Harris County. To quantify and measure community recovery at the same spatial scale (CBG level), we examine human mobility patterns. As a proxy of disaster recovery, human mobility has been shown to reflect the functioning of communities [8-10]; many studies have used human mobility data to gauge the impact of disasters on community functionality and recovery [9, 11, 12]. Based on the specification of the extent of vulnerability and the speed of recovery across all spatial areas, the analysis focuses on the local and spillover effects of their relationships. Spatial spillover effects exist in a variety of spatial processes, such as the spread of environmental harm [13], human behaviors [14], innovations [15], perceptions [16], and technologies [17] across cities or different areas of a city. Spatial processes also play an important role in community resilience to hazards [1, 18]. Yet, the characterization of spatial diffusion processes and spatial spillover effects shaping community resilience are the object of slight attention in the literature.

Recognizing these gaps, this study implements spatial regression models to identify the spatial spillover effect, spatial reach, and spatial decay effects of residential flood damage on community recovery. In particular, the analysis focuses on answering the following research questions: (1) To what extent does the severity of residential damage explain variations in the speed of recovery? (2) to what extent do residential damage result in spatial spillover effects on surrounding areas? (3) How extensive is the spatial reach of these spillover effects? and (4) Are the spillover effects are homogenous across all areas in the affected region? The findings reveal important characteristics related to the spatial diffusion of residential damage and their effects on community recovery. First, the findings show a significant relationship between the extent of residential damage and local community recovery speed. Areas with a greater extent of residential damage have a slower recovery speed. Second, the extent of this statistical relationship is not locally contained but spills over to surrounding areas. In other words, the extent of residential damage in one area influences the recovery speed of the surrounding areas with a spatial reach of up to 31.2 miles. Third, while the effect of the spatial reach of residential damage on community recovery speed diminishes with distance from the highly damaged areas, it is not homogenous across all areas. Fourth, areas with a greater concentration of points of interest (POIs) and road networks show less sensitivity to the residential damage in other farther areas. These findings offer a novel data-driven characterization of the spatial diffusion of residential flood damage effects on community recovery and move us closer to a better understanding of complex spatial processes that shape community resilience to hazards. Figure 1 depicts an overview of our study framework.

In sum, the novel aspects and contributions of this study are threefold. First, the study uncovers important characteristics of the spatial diffusion process related to built-environment vulnerability effects on community recovery. The findings contribute to a better understanding of the characteristics of spatial diffusion processes that extend the reach of built-environment vulnerability beyond local effects to regional recovery speed. Also, the discovery of the heterogenous spillover effect and its sensitivity to features of urban structure inform about ways to dissipate the spillover effects of built-environment vulnerability through urban planning strategies that influence urban structure. This scientific contribution informs interdisciplinary fields of civil engineering, urban science, geography, and disaster science about the complex spatial processes that shape community resilience. Second, the study harnesses unique high-resolution datasets to quantify and measure residential damage extent and community recovery. The flood claim dataset used in this study involves both National Flood Insurance Program (NFIP) and the Individual Assistance (IA) program, which capture both insured and uninsured losses and provide a more complete picture of residential damage than NFIP data alone, which excludes uninsured losses (leading to imprecise assessments in events such as Hurricane Harvey in which a great extent of damage occurred to uninsured properties). Also, the fine-grained human mobility data used in this study enable reliable quantification of community recovery speed, since human mobility is a reliable proxy for capturing the functioning of communities based on population activities. The assessment of these high-resolution datasets together in a unique event (Hurricane Harvey) provides novel insights regarding the spatial patterns of residential damage and community recovery at a scale and resolution never attempted before. Third, the findings provide valuable perspectives into non-local effects of residential damage and spatial spillover effects for emergency managers and public officials to better evaluate the benefits of flood risk reduction measures for the built environment for enhancing community resilience.

The rest of this paper is organized as follows. Section 2 introduces the study area and datasets. Section 3 introduces variables and models. Section 4 presents the experimental results. Finally, discussion and concluding remarks are presented in section 5.

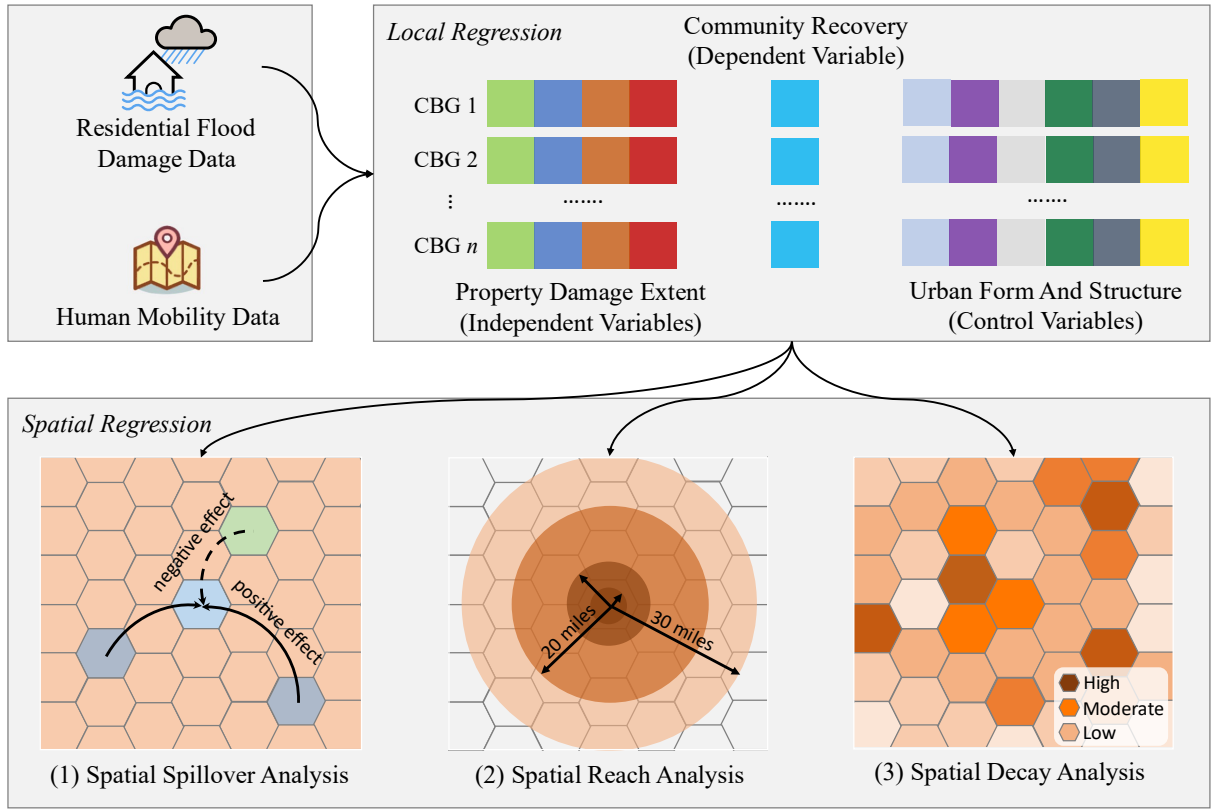

Figure 1. Conceptual framework for examining the relationship between residential flood damage and community recovery speed.

## 2. Study context and datasets

### 2.1 Hurricane Harvey in Harris County

This study collected and analyzed data from Harris County, Texas, which includes the Houston metropolitan area, one of the regions most adversely affected by the 2017 Hurricane Harvey. On August 25, 2017, Hurricane Harvey, a devastating Category 4 hurricane, made landfall, inundating Harris County with record level rainfall. Large areas of the Houston downtown and some western areas of Harris County were subsequently flooded [19]. Also, due to the release of water from Barker and Addicks Reservoir, the western part of Harris County experienced extensive and prolonged flooding [20]. The impacts of Hurricane Harvey continued until September 1, 2017, when it weakened, and residents started to recover from the storm impacts. The magnitude of flood impacts of Harvey in the Houston metropolitan area, one of the largest metropolitan areas in the U.S., make the region and the hazard event a unique testbed for examining the characteristics of spatial diffusion of built-environment vulnerability effects on community recovery. We collected data of Harris County in 2,144 CBGs from the period between August 1, 2017, to September 30, 2017.

### 2.2 Flood damage claims dataset

To evaluate residential damage within communities after flood events caused by Hurricane Harvey, we relied on flood damage claim data as a key indicator of property damage extent. The data, sourced from both the NFIP and IA program, capture both insured and uninsured losses and provide a more comprehensive approach to quantifying built-environment vulnerability (previous studies using NFIP data only could capture insured losses, providing an incomplete picture of vulnerability extent [21-23]). The NFIP provides policies to cover flood-induced damage to structures and their contents, unlike typical homeowners' insurance, and is obligatory for properties in the Federal Emergency Management Agency (FEMA)-defined 100-year floodplain with federally backed mortgages [24]. In contrast, the IA program is triggered only by presidential disaster declaration. Individual Assistance offers financial help to those affected by disasters, covering essential uninsured expenses to meet basic needs and support the recovery process without intending to replace insurance or fully compensate losses [25]. To ensure a comprehensive assessment, we augmented NFIP flood claim records by integrating record from IA program, allowing for capturing both insured and uninsured losses.

Our analysis focuses primarily on building structures, excluding damage to contents. We utilized building claim payments from NFIP and real property damage assessments by FEMA in the IA to gauge property damage. To maintain accuracy, only records with non-zero damage values from both NFIP and IA datasets were included, as zero values may indicate damage unrelated to flood hazards.

### 2.3 CBG-to-CBG human movement dataset

We used the high-resolution human mobility dataset collected from Spectus Inc. to identify CBG-to-CBG human movement. Spectus' location intelligence platform collects mobility data of anonymized devices of users who have opted in to provide access to their location data for research purposes through a CCPA (California Consumer Privacy Act)-compliant and GDPR (General Data Protection Regulation)-compliant framework [26].

In this study, people who left their home CBGs and dwelled in the other CBGs were viewed as CBG-to-CBG human movement; thus, the movement rate indicates that the percentage of people who left their home CBGs. To this end, the first step of data processing was to use Spectus data to identify home CBG of devices. We adopted the dwell time of devices in the Device_Location table, which refers to the duration of the visit, as an indicator for detecting home CBG. If a device stopped in a CBG for more than 24 hours, the CBG would be recognized as its home CBG. Then, Table Stop was used to extract which CBGs the users have visited. The indicator dwell time records the stop duration of devices and was adopted to decide whether it was a visit to a CBG or just a stop-by. If the duration of time an anonymized device dwelled in one CBG exceeded the threshold (4 hours in this study), then a visit to that CBG was recorded. Through the above steps, the dataset, including CBG-to-CBG human movement, was obtained. The original data was hourly and was temporally aggregated to daily.

## 3. Methodology

### 3.1 Variables

#### 3.1.1 Independent variables

We calculated four metrics from the flood damage claim datasets as independent variables to represent residential damage extent. Initially, we collected the 2017 property value dataset for Harris County through the Texas Natural Resources Information System [27]. This dataset includes an entry labeled "market value," which represents the estimated current value of the entire property, considering both the land and improvements. The dataset takes into account the combined worth of the raw land and any structures on it in the current real estate market.

Since the NFIP and IA datasets are geospatially point-based and the property value dataset is geospatially polygon-based, to ensure accurate spatial alignment and facilitate a comprehensive assessment of property damage, we integrated the NFIP and IA datasets into the property value dataset using geographical analysis techniques. We used the shortest Euclidean distance method to accurately map claim points to specific building polygons in the property value dataset. Consequently, for any given property in Harris County, we could ascertain both the NFIP or IA claim value and the property's market value for that year. Given the higher amount scale typically associated with NFIP claim values as compared to IA claim values, for properties with records from both NFIP and IA, we exclusively used NFIP claim data. For other properties, a linear regression method was implemented to establish a quantitative linear relationship between the two datasets, as detailed in Figure S1 in the Supplementary Information. This step normalized the NFIP and IA data to the same scale. Subsequently, for each property, we calculated the property damage extent (PDE) as follows:

$$PDE_{ij} = \frac{CV_{ij}}{MV_{ij}} \tag{1}$$

where $PDE_{ij}$, $CV_{ij}$, and $MV_{ij}$ denote the PDE, claim value and market value, respectively, of the property $i$ in the CBG $j$. A total of 72,755 PDE records from Harris County were computed. This data was further analyzed to derive four key statistical measures of PDE at the CBG level.

- *Number of claims*. This frequency metric focuses on counts related to the occurrence of claims in a CBG.
- *Mean of PDE*. This central tendency metric describes the average of PDE values in a CBG.
- *Standard deviation of PDE*. This dispersion metric indicates the variability of PDE values in a CBG.
- *Major damage level of PDE*. This metric captures the number of properties whose PDE is larger than 0.5 in a CBG, providing insight into the severity of the flooding events.

### 3.1.2 Dependent variable

We calculated the recovery rate, which represents functionality restoration speed of CBGs, using the movement rates retrieved from CBG-to-CBG human movement dataset. We first established baselines based on normal visit patterns. Since the hurricane made landfall on August 25, 2017, the CBG-to-CBG movement data from August 1 to August 21 served as a baseline when no disturbance occurred to the residents' normal lifestyle [9]. Since weekday and weekend visit patterns differed, the baseline period was calculated as a weekly pattern, considering each day as a unit [8]. Each CBG's daily visits to CBG were compared to the corresponding baselines to compute the percentage change using Equation (2):

$$PC_{i,t} = \frac{MR_{i,t} - BL_{i,t}}{BL_{i,t}} \tag{2}$$

where $PC_{i,t}$ is the percentage change of visits to CBGs from home CBG $i$ on date $t$; $MR_{i,t}$ is the movement rate of CBG $i$ on date $t$, and $BL_{i,t}$ is the calculated baseline value of movement rate of CBG $i$ corresponding to the date $t$.

Due to the impact of the hurricane, the CBG-to-CBG movement rate decreased as the population stayed at their homes to avoid harm then rebounded to the normal situation as the hurricane weakened and left the region [9]. Therefore, the recovery extent and recovery duration are critical milestones of community recovery. We conceptually described the trend of percent change of movement rates $PC_{i,t}$ over time in Figure S2 in the Supplementary Information. In this trend, we identified the duration for return to a new steady state in the period after the maximum change of movement rate. Since a new steady state in the aftermath of the hurricane can vary compared to the pre-disaster status, we defined the new steady state as having no substantial difference in terms of the percent changes of movement rate. That is, in this study, the difference in the percent change between two consecutive days is within a threshold of 10%. The recovery extent is the difference between the maximum change and 90% of the mean baseline. The recovery rate $RR$ is then calculated using the Equation (3):

$$RR_i = \frac{PC_{i,t_s} - PC_{i,t_n}}{t_s - t_n} \tag{3}$$

where $RR_i$ is the recovery rate of CBG $i$; $t_s$ represents the date when the movement rate was at a minimum; $t_n$ represents the date when the movement rate recovered to the new steady state. The term $t_s - t_n$ represents recovery duration and the term $PC_{j,t_s} - PC_{j,t_n}$ represents recovery extent.

### 3.1.3 Control variables

To further evaluate factors that explain variations in the extent of spillover effect of residential flood damage on community recovery in different areas, six control variables were selected: (1) population density, calculated by dividing the total population of the CBG by its land area; (2) minority segregation, indicating the extent to which two racial groups are evenly distributed across CBGs; (3) income segregation, indicating the extent to which two income groups are evenly distributed across CBGs; (4) human mobility index, indicating the level of human mobility and activity; (5) POI density, capturing the distribution of physical facilities; (6) road density, capturing the distribution of transportation network. The details in terms of these control variables can be found in Note S1 in the Supplementary Information. Considering the different scales among values of these variables, we used min-max scaling to scale all variables to [0, 1] for the further process. The descriptive statistics of variables in this study can be found in Table 1.

Table 1. Descriptive statistics of variables

| Variable Type | Variables | Abbreviation | Min | Max | Mean | Std Dev |
|---|---|---|---|---|---|---|
| Dependent Variable | Recovery Rate | RR | 0.000 | 0.877 | 0.188 | 0.134 |
| Independent Variables | Number of claims | NC | 0 | 707 | 34 | 61.991 |
| | Mean of PDE | MP | 0.000 | 0.925 | 0.288 | 0.181 |
| | Standard deviation of PDE | SDP | 0.000 | 0.588 | 0.155 | 0.101 |
| | Major damage level of PDE | MDP | 0 | 492 | 24 | 39.566 |
| Control Variables | Population density | POP | 0.000 | 0.021 | 0.002 | 0.002 |
| | Minority segregation | MS | 0.000 | 4.647 | 0.351 | 0.484 |
| | Income segregation | IS | 0.000 | 0.002 | 0.000 | 0.000 |
| | Human mobility index | HMI | 6729 | 3561566 | 96664 | 122700 |
| | POI density | POI | 0.000 | 0.006 | 0.000 | 0.000 |
| | Road density | RD | 0.000 | 0.034 | 0.011 | 0.005 |

## 3.2 Methods

### 3.2.1 Ordinary least squares regression model

To capture the relationship between residential flood damage extent metrics and community recovery speed and to understand the relative importance of each metric, we employed an ordinary least squares (OLS) regression model in Equation (4). This model also served as the benchmark model to provide a foundational comparison for assessing the effectiveness of spatial regression models in capturing these relationships.

$$Y = \beta_0 + X_1\beta_1 + X_2\beta_2 + \varepsilon \tag{4}$$

where $Y$ is the recovery rate of CBGs; $X_1$ and $X_2$ are matrixes of independent variables and control variables, respectively; $\beta_0$ is the constant term; $\beta_1$ and $\beta_2$ are vectors of coefficients; $\varepsilon$ is the vector of random errors. Pearson's correlation test was then conducted for the correlation analyses to examine statistical significance and determine feature importance.

### 3.2.2 Spatial lag of X model

We conducted a spatial autocorrelation test before specifying spatial regression models [28]. We applied the Global Moran's I test to see if there are spatial interrelationships between neighboring CBGs. The null hypothesis for the Global Moran's I test holds that the features being analyzed are randomly distributed in the study area. Therefore, a significant $p$-value indicates the spatial patterns are not random [29]. To define the neighbors and calculate the Global Moran's I, we first constructed an inverse distance spatial weight matrix $\omega_{ij}$ to assign weights to CBGs according to geographical distance (Equation 5). The Global Moran's I was then calculated by Equation (6). The Global Moran's I generally ranges from -1 to 1. If the value is significantly greater (less) than 0, there are globally positive (negative) correlations between geographically close CBGs [29].

$$\omega_{ij} = \begin{cases} \dfrac{1}{d_{ij}}, i \neq j \\ 0, i = j \end{cases} \tag{5}$$

where $\omega_{ij}$ is a row normalized spatial weight matrix of $n \times n$ that defines the geographical relations between CBG $i$ and CBG $j$; $d_{ij}$ is the spatial Euclidean distance between the centroid of CBG $i$ and CBG $j$.

$$I_G = \frac{n \sum_{i=1}^{n} \sum_{j=1}^{n} w_{ij}(x_i - \bar{x})(x_j - \bar{x})}{\sum_{i=1}^{n} \sum_{j=1}^{n} w_{ij} \sum_{i=1}^{n}(x_i - \bar{x})^2} \tag{6}$$

where $I_G$ refers to the Global Moran's I across all CBGs; $n$ refers to the number of CBGs; $x_i$ and $x_j$ refer to independent variables (e.g., number of claims) in CBG $i$ and CBG $j$; $\bar{x}$ refers to the average value of independent variables.

One of the key hypotheses in this study is that the recovery rate of a given CBG is affected not only by its own residential damage but also by that of adjacent CBGs. The spatial lag of X (SLX) model is suitable for investigating how independent variables in the local area and neighboring areas are potentially interrelated with dependent variables in the local area (Equation 7) [30]. This model is distinctive in its ability to address not just correlations within a local vicinity but also to incorporate spatial dependencies, distinguishing it from other spatial regression models, such as spatial lag, spatial Durbin, and spatial error models, which may not simultaneously account for both non-spatial and spatial effects in this study [30].

$$Y = \beta_0 + \rho WX + \beta X + \varepsilon \tag{7}$$

where $Y$ is the recovery rate of CBGs; $X$ is the matrix of independent variables and control variables; $\rho$ is the coefficient for the spatial lag term; $W$ is the spatial weight matrix; $\beta_0$ is the constant term; $\beta$ is the vector of coefficients; $\varepsilon$ is the vector of random errors.

Then, to estimate spatial reach of spatial spillover effects, we set a threshold $D$ within the spatial weight matrix $\omega_{ij}$ (Equation 8). This threshold serves as a cutoff point: if the distance between two CBGs is less than $D$, then the spatial weight is applied between them; if it is greater, the spatial weight is set to zero. Considering the furthest distance between CBGs in Harris County is around 70 miles, we established a series of incremental thresholds at 0.1-mile intervals from 0 to 70 miles. In this manner, we can incrementally test and identify the maximum distance at which the spatial spillover effect is observable. This approach allows for a precise estimation of the spatial reach of the spillover effects in the context of CBG interactions.

$$\omega_{ij} = \begin{cases} \dfrac{1}{d_{ij}}, i \neq j \text{ and } d_{ij} \leq D \\ 0, others \end{cases} \tag{8}$$

where $D$ ranges from 0 to 70, incrementing in steps of 0.1.

### 3.2.3 Spatial decay model

In addition to investigating the presence and maximum reach of spatial spillover effect, we also explored the rate at which the effect diminishes over space. We developed a spatial decay (SD) model based on Rachlin's hyperbolic delay discounting model [31]. Rachlin's model, typically used to describe how individuals value rewards over time, proposes that the perceived value of a reward diminishes at a decreasing rate as the delay increases. Unlike a constant rate of decrease per time unit, hyperbolic discounting depicts a decline that slows over time [31]. The appropriateness of the Rachlin's model for applications in natural disaster contexts is supported by its use in developing models of human mobility during large-scale crises [32]. Adapting this concept, our adapted spatial decay model to examine the decay of spatial spillover effect in the context of the hurricane is as follows:

$$k_i = \frac{RR_0 - RR_i}{RR_i \sum_{i,j=1}^{n} \omega_{ij} PDE_i} \tag{9}$$

where, $RR_i$ is the recovery rate of CBG $i$; $PDE_i$ is the property damage extent feature of CBG $i$, here we use the number of claims; the spatial decay coefficient $k_i$ measures the degree of the decay in CBG $i$, with a larger value corresponding to more rapid decay; $\omega_{ij}$ is determined by the distance threshold $D$ obtained through spatial reach analysis. $RR_0$ represents the ideal recovery rate, here we use the maximum recovery rate among all the CBGs.

## 4. Results

### 4.1 Descriptive analysis of variables

We first present an overview of the descriptive statistics and geographical patterns of our variables. Figure 2a displays the frequency distribution of PDE values, which vary from 0 to 1. The average PDE for Harris County is 0.3902, indicating that, on average, 39.02% of the economic value of a property was damaged due to flooding. A significant concentration of PDE values falls between 0.2 and 0.4, indicating the majority of property damage is moderate. Figure 2b illustrates the high-resolution spatial distribution of the residential damage at the individual property level with the inclusion of the FEMA 100-year floodplain and major river streams. This map visualizes the geographic relationship between residential flood damage and high flood-risk areas. Two notable clusters of properties with high PDE values are located in the northeastern and southeastern areas of downtown Houston. The serious damage in the northeast could be attributed to its higher population density and older infrastructure. Conversely, the concentration in the southeast may be linked to its geographical characteristics as a coastal inlet and low-lying hurricane impact zone. In addition, the map highlights significant property damage flanking the FEMA 100-year floodplain and adjacent to river streams. Figure 2c shows the number of claims at the CBG level, with an average of 33.93 claims per CBG. This figure supports the observed patterns in the southeast and northeast areas of downtown Houston, where a relationship between the number of claims and the severity of property damage is apparent. Here, properties are not only more severely damaged but also experience a broader scope of damage. In contrast, the northwestern Houston downtown area, despite having a lower severity of property damage, exhibits a high number of claims in CBGs. This indicates that while individual property damage may be less severe, it is more broadly distributed across the area. Figure 2d provides a spatial distribution of recovery rate at the CBG level. From these three spatial distribution figures, we conclude that areas with more extensive residential flood damage and a larger number of claims, such as the northeastern downtown area, tend to have a slower recovery rate. Conversely, areas with lesser property damage and fewer claims, exemplified by the southwestern downtown area, experience a quicker recovery. This trend highlights a spatial correlation between the recovery rate and the metrics of property damage, which signals the need for further spatial regression analysis. To offer a more comprehensive view of the situation, we have also included spatial distribution for other independent and control variables (Figure S3 to Figure S11 in the Supplementary Information).

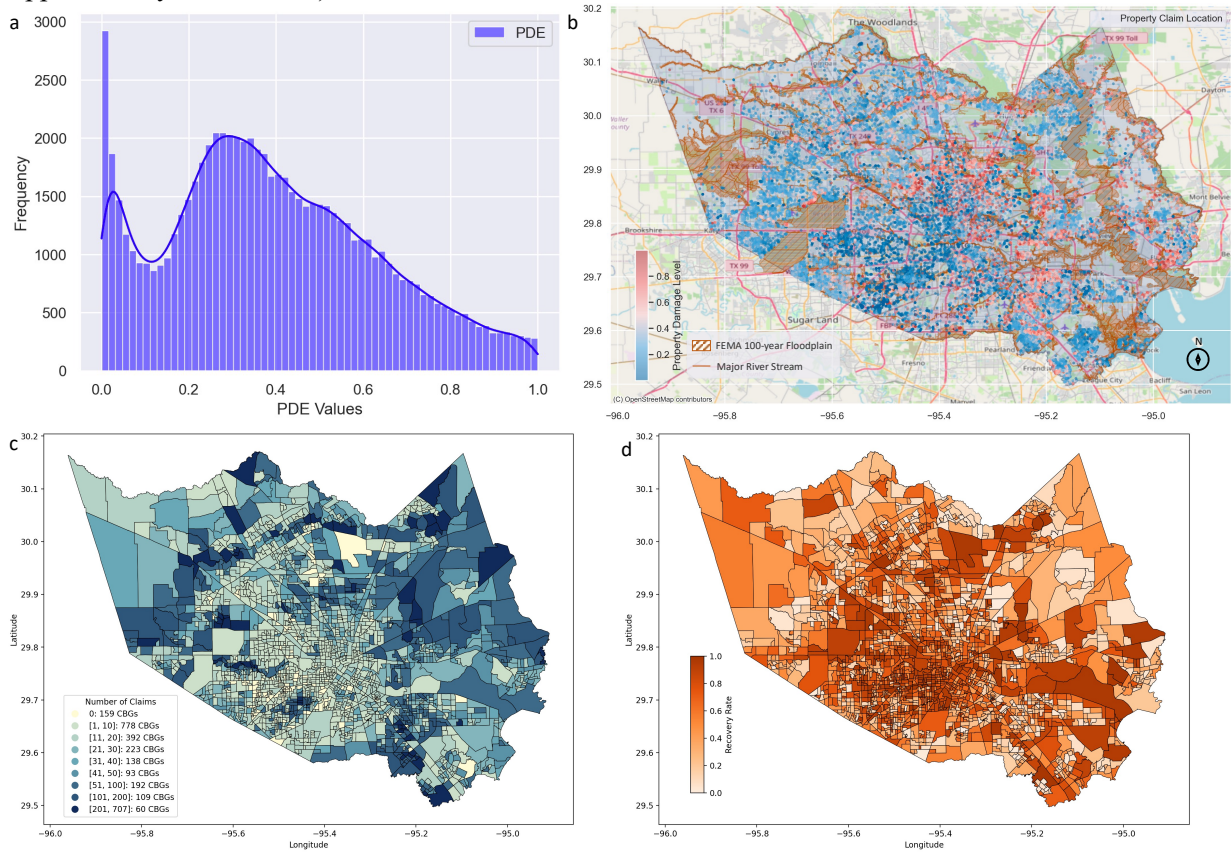

Figure 2. (a) Frequency distribution of PDE values. (b) Spatial distribution of property damage locations and corresponding PDE values. (c) Spatial distribution of number of claims at the CBG level; (d) Spatial distribution of recovery rate at the CBG level.

## 4.2 Local driving factors of community recovery

It is crucial to address the potential issue of multicollinearity among the independent variables in relation to the dependent variable before correlation analysis. This study employed the variance inflation factor (VIF) method to examine the multicollinearity degree. According to the standards [33], if the VIF of an independent variable is greater than or equal to 10, the regression model is assumed to have serious multicollinearity, and conversely, there is no significant multicollinearity problem. In this case, all VIFs of our variables are below 5.0000 (Figure 3b), so there is no significant multicollinearity among our variables; thus multicollinearity would not affect the model estimation accuracy.

Figure 3a presents the Pearson's correlation between our independent variables and dependent variable. All property damage extent features (e.g., NC, MP, SDP, and MDP) exhibit negative correlations with recovery rate at the significant level of 0.001. This indicates that greater extents of property damage are associated with slower recovery rate at the CBG level in Harris County post-Hurricane Harvey. Additional tests of multicollinearity and correlation for the control variables are depicted in Figures S12 and S13 in the Supplementary Information. All six control variables passed the multicollinearity test. The POI density and human mobility index demonstrate a significant positive correlation with the recovery rate, whereas income segregation displays a notable negative correlation. The relationships between the other control variables and the recovery rate are statistically insignificant.

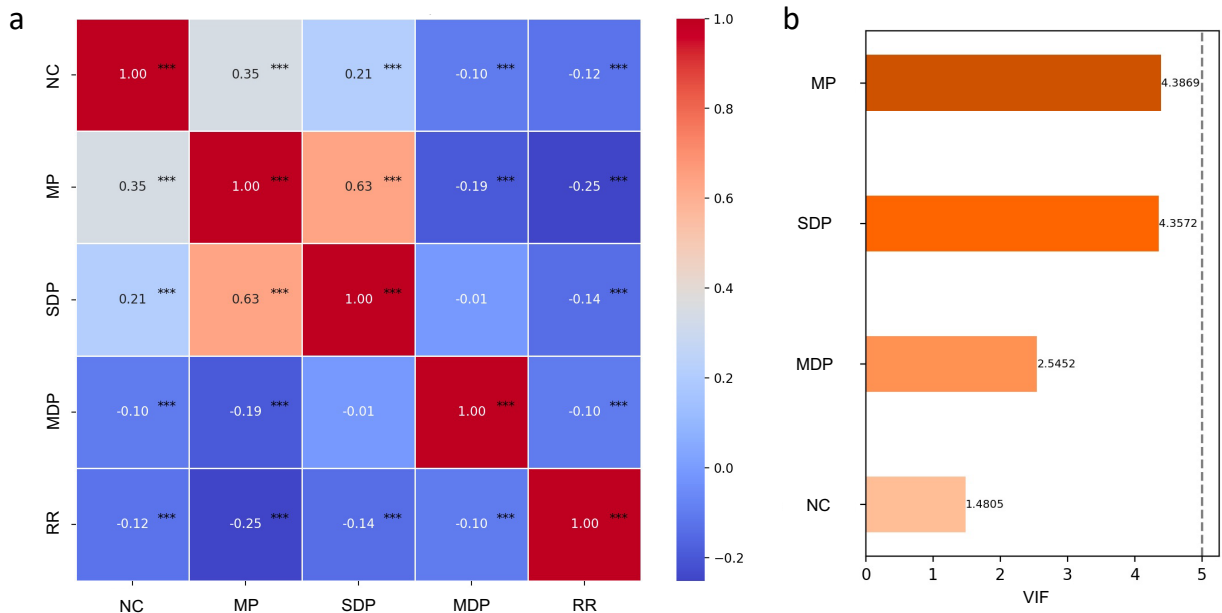

Figure 3. (a) Pearson correlation heatmap for the property damage extent features. The number within the cell indicates the Pearson correlation coefficient. ***$p<0.001$, **$p<0.01$, *$p<0.05$. (2) Multicollinearity test results of the property damage extent features using Variance Inflation factor (VIF).

Figure 4 presents the scatter plots and linear regression lines with $R^2$ between recovery rate and property damage extent features. In regression models, $R^2$ serves as an indicator of the quality of model fit, with higher values indicating better fit [34]. Notably, the correlation between mean of PDE and the recovery rate yields the highest $R^2$ of 0.6331, indicating that 63.31% of the variation in recovery rate can be explained by mean of PDE, while the other three $R^2$ of property damage extent features are relatively low. This could be because mean of PDE captures the central tendency of probabilities which is likely a critical factor in determining recovery rates. For the six control variables, we similarly conducted an $R^2$ analysis in Figure S15 of Supplementary Information.

In summary, the four property flood damage extent features (NC, MP, SDP, and MDP) proposed in this study serve as robust indicators of local community recovery speed. The findings support the notion that the severity of residential damage correlates inversely with the local community recovery rate. The more substantial and widespread the residential property damage, the slower the community's recovery. This result confirms the relationship between residential flood damage and recovery speed locally (i.e., for each CBG). In the next step, the potential for spatial correlation and its influence across a broader geographic area are further investigated.

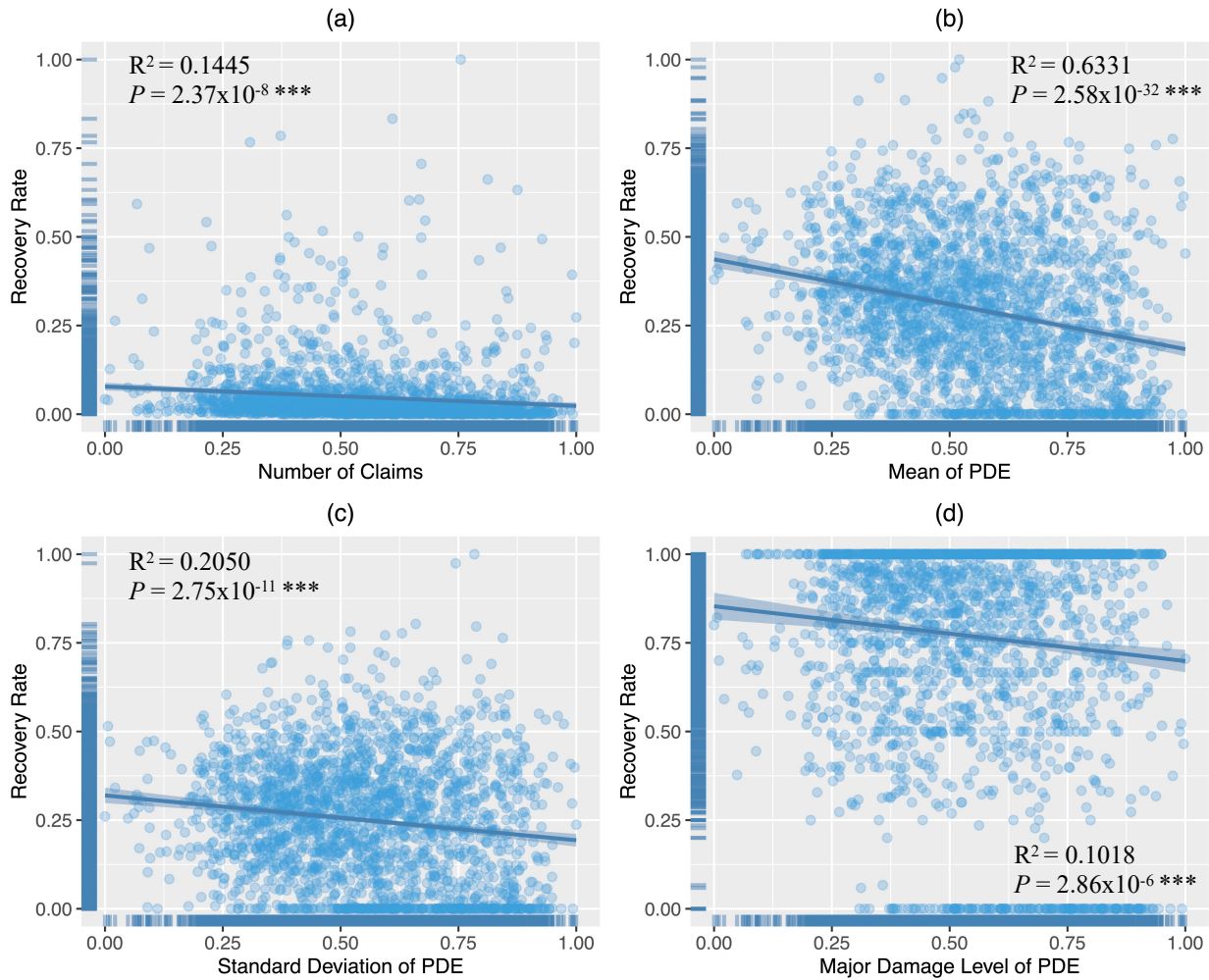

Figure 4: Correlation between property damage extent features and recovery rate. (a) Number of claims. (b) Mean of PDE. (c) Standard deviation of PDE. (d) Major damage level of PDE. The variable distribution was given by rugs. We used the OLS regression model to fit the data and the confidence interval was set to be 90%. All measures are statistically significant with ***$p<0.001$.

### 4.3 Spatial spillover analysis

To investigate the spatial spillover effect between residential damage and community recovery speed, we began by conducting the Global Moran's I test. The test yielded the Global Moran's I value of 0.172 with a

*p*-value of 0.001, indicating a globally positive correlation among neighboring CBGs. As a result, the SLX model was used to assess the spatial spillover effect further. For comparison, and to assess the robustness of our regression findings, the results and performance from the OLS model were also included as a reference point in Table 2.

To evaluate the performance between OLS and SLX models, we first used $R^2$ and adjusted $R^2$. The SLX model demonstrates higher values with $R^2$ of 0.335 and adjusted $R^2$ of 0.289, compared to the OLS model with an $R^2$ of 0.207 and adjusted $R^2$ of 0.203. This result suggests that the spatial regression model is able to account for a greater extent of variation in the recovery rate than the local regression model. Furthermore, the Akaike information criterion (AIC) serves as an indicator of the prediction error of the model, where a lower value denotes a better model fit [34]. The AIC for the SLX model is -1735.931, which is lower than the OLS model's AIC of -1425.000, indicating a more precise fit for the SLX model. Lastly, the log-likelihood statistic measures the model's ability to explain the observed data [34]. Here, the SLX model achieves a log-likelihood of 889.965, surpassing the OLS model's log-likelihood of 723.360. This higher log-likelihood signifies that the SLX model fits the observed data more accurately than the OLS model.

In the SLX model, the regression coefficients for NC, MP, and MDP were all significantly negative at the 5% significance level. This result indicates that residential flood damage not only hinders local community recovery but also exerts a notable suppressive effect on the recovery speed of geographically interconnected CBGs, thus validating the presence of spillover effect. Control variables like POP and MS did not show statistical significance; however, the coefficients for RD and IS were notably negative (-0.194* and -0.182***, respectively), suggesting that increases in RD and IS are detrimental to recovery rate both locally and regionally. In contrast, POI and HMI presented positive coefficients (0.155*** and 0.211***, respectively), indicating their role in expediting recovery. Notably, RD, which was not significant in the OLS model (0.003), became significant in the SLX model (-0.194*), highlighting its relevance in post-disaster recovery processes, particularly in surrounding areas. Furthermore, it is observed that the absolute values of the regression coefficients are greater in the SLX model (except for SDP in SLX and POP, RD, and MP in both models). This result suggests that the SLX model's coefficients better account for the combined effects of local and spatial factors while the OLS model fails to capture them. In other words, the effect on community recovery may be underestimated if the spatial spillover effect is not considered.

While it has been established that the independent/control variables can either inhibit or promote the recovery rate both locally and regionally, quantifying the extent to which these variables affect recovery processes is fundamental to devising effective interventions and policies for post-disaster rehabilitation. LeSage and Pace (2021) pointed out that analyzing the spatial spillover effects between regions directly based on the regression coefficients of the spatial interaction terms may lead to erroneous conclusions [35]. Therefore, it is important to decompose the total effect into direct effect (local effect) and indirect effect (spatial-spillover effect). The direct effect explains how a change in the explanatory variable for a unit influences the dependent variable in the local unit. The indirect effect quantifies the impacts on the dependent variable as it manifests itself in other neighboring units. Formally, the direct effect plus the indirect effect equals the total effect [35]. Table 2 illustrates that for all four independent variables, the absolute values of the indirect effect surpass those of the direct effect, although SDP shows no significance. This result indicates that the examined property damage extent features exert a more pronounced dampening impact on neighboring CBGs than on the CBGs themselves. Among these, NC exhibits the greatest inhibitory effect both locally and regionally, followed by MP. This pattern suggests that in the context of community recovery, the scale of residential damage (measured by number of claims) is more important than the severity of individual property damage (measured by mean of PDE), particularly in terms of spillover effects. These findings underscore the importance of addressing the breadth of residential damage in hurricane recovery strategies, not merely the areas with the most severe damage.

Furthermore, to validate the robustness of our findings, we conducted a series of robustness tests by altering the spatial weight matrix using methods of shared boundary, k nearest neighbors, and inverse square of geographical distance. Subsequent recalculations, detailed in Table S1 of the Supplementary Information, revealed consistency with our primary results and model performance. This consistency across various methods of spatial weighting substantiates the reliability and stability of our approach.

Table 2. Regression result and model performance of OLS and SLX model

| Variables | OLS Model | SLX Model (Global Moran's I: 0.172, *P* value: 0.001***) | | | |
|---|---|---|---|---|---|
| | Coefficient | Coefficient | Direct effect | Indirect effect | Total effect |
| *Independent Variables:* | | | | | |
| NC | -0.173*** | -0.319*** | -0.199*** | -0.375*** | -0.573*** |
| MP | -0.268*** | -0.278** | -0.124*** | -0.234** | -0.358** |
| SDP | -0.187*** | -0.010 | -0.001 | -0.002 | -0.003 |
| MDP | -0.084*** | -0.089*** | -0.021* | -0.021** | -0.022* |
| *Control Variables:* | | | | | |
| POP | -0.158 | -0.079 | -0.096 | -0.051 | -0.147 |
| RD | 0.003 | -0.194* | -0.162* | -0.306*** | -0.468** |
| POI | 0.133*** | 0.155*** | 0.318*** | 0.221*** | 0.537*** |
| MS | -0.142 | -0.236 | -0.007 | -0.004 | -0.010 |
| IS | -0.121*** | -0.182*** | -0.091*** | -0.171** | -0.262*** |
| HMI | 0.196*** | 0.211*** | 0.179* | 0.307** | 0.485*** |
| *Model Performance* | | | | | |
| $R^2$ | 0.207 | 0.335 | | | |
| Adjusted $R^2$ | 0.203 | 0.289 | | | |
| Log-likelihood | 723.360 | 889.965 | | | |
| AIC | -1425.000 | -1735.931 | | | |

Note: Number. of observations: 2, 144 CBGs; ***, **, and * refer to the significance level at 1%, 5%, and 10%, respectively.

### 4.4 Spatial reach analysis

The above analyses confirm the existence of spatial spillover effects of residential damage on community recovery speed. Yet, the specific extent of this spatial reach remains unknown. Quantifying the spatial reach is crucial for deepening our understanding of the impact of residential damage and informing practical disaster recovery planning. We first defined distance thresholds within the spatial weight matrix and then incrementally ran the SLX model using intervals of 0.1 miles to analyze the direct, indirect, and total effects at each specified spatial distance. The results of NC and POI density are presented in Figure 5. The other plots for the independent and control variables are shown in Figure S15 to Figure S19 in Supplementary Information. It should be noted that we plotted only the significant independent and control variables in Table 2.

In Figure 5, the direct effects remain stable as this encapsulates only the local impact devoid of spatial considerations. Conversely, the indirect and total effects require examination across varying distances. For NC (as well as MP, MDP, RD, and IS in the Supplementary Information), negative spatial spillover effects were observed within a significant cut-off range (61.8 miles), which nearly spanned the maximum distance between any two CBGs in Harris County. This implies that NC consistently exerts an inhibitory influence on the recovery of neighboring CBGs across the county. The spatial spillover effects exhibit a nonlinear pattern, intensifying up to a minimum distance (31.2 miles) before diminishing. This increase suggests that as more surrounding CBGs fall within the expanding distance threshold, they collectively exert a stronger inhibitory effect on the recovery speed of a single CBG. Beyond 31.2 miles, the inhibitory effect of spatial spillover gradually diminished, indicating that although the number of encompassing neighboring CBGs further increases, the influence of the variable of interest on the one CBG decreases with increasing distance. This phenomenon verifies the concept of spatial reach: the effect of a variable or intervention is bound by a spatial limit, exerting its strongest influence within a specific radius. Beyond this range, its ability to affect outcomes decreases [36]. The NC plot demonstrates that after reaching a minimum, the effects begin to decrease, which suggests that the effect is most potent at an intermediate distance (not too proximate nor too remote from the local CBG). Accordingly, we specify the spatial reach for the effect of residential damage on community recovery speed to be 31.2 miles. For POI and HMI, the spatial spillover effect is initially positive, increasing to a point before declining. Their maximum distance threshold concurs with that determined for NC, leading to similar conclusions regarding their spatial impact on recovery.

To assess the robustness of our method for estimating spatial reach, we also employed the above three distinct spatial weights to determine the cut-off and min/max distances. The results, as presented in Table S1 of the Supplementary Information, show that these distances varied within a ±20% deviation from those obtained through the approach in the main text. Such results verified the reliability of our approach in calculating the spatial reach.

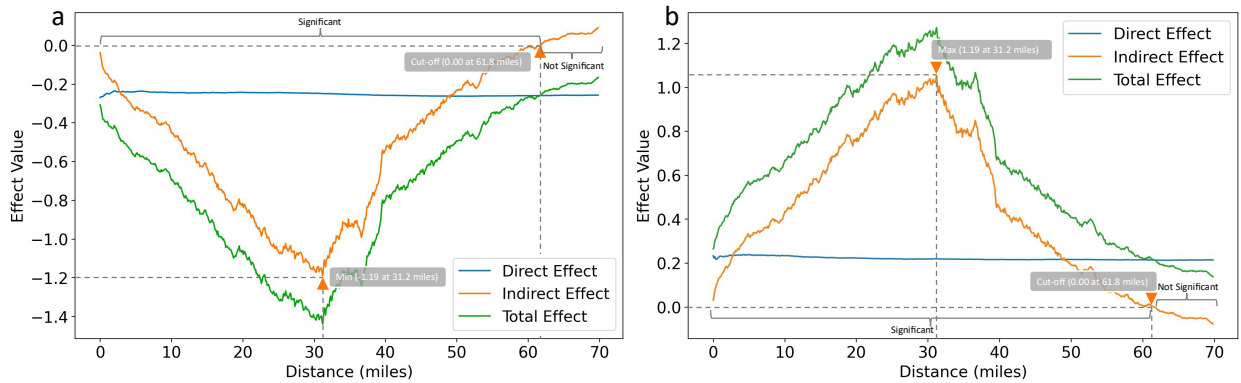

Figure 5. Decomposition effects of the independent and control variables as the distance threshold changes in the spatial weight matrix. (a) Independent variable of number of claims. (b) Control variable of POI density. The min/max arrows indicate the distance at which the indirect effect reaches its minimum/maximum value. The cut-off arrows denote the distance within which the decomposition effect is at least 10% significant; effects beyond this point are not significant.

### 4.5 Spatial decay analysis

The previous analyses confirmed the existence of spatial spillover effects and spatial reach of residential flood damage on community recovery rate. Given that spatial spillover effects are not uniform across all CBGs, however, a more granular examination is necessary to accurately quantify the heterogeneity of spillover effects across CBGs. To this end, we adopted a spatial decay model to determine the spatial decay coefficient for each CBG. The coefficient indicates the rate at which spatial spillover effect wanes with distance where a higher value signifies a more rapid decrease in effect. This means that for CBGs with high spatial decay coefficients, the recovery of the community is more sensitive to the influence of proximate neighbors as opposed to those further away. From Figure 6, CBGs with high spatial decay coefficients are predominantly located in downtown Houston and its southwestern areas. In contrast, CBGs in other areas exhibit relatively lower spatial decay coefficients. This distinction suggests that the recovery rates of CBGs in the downtown and southwestern areas are more tightly affected by their immediate neighboring CBGs.

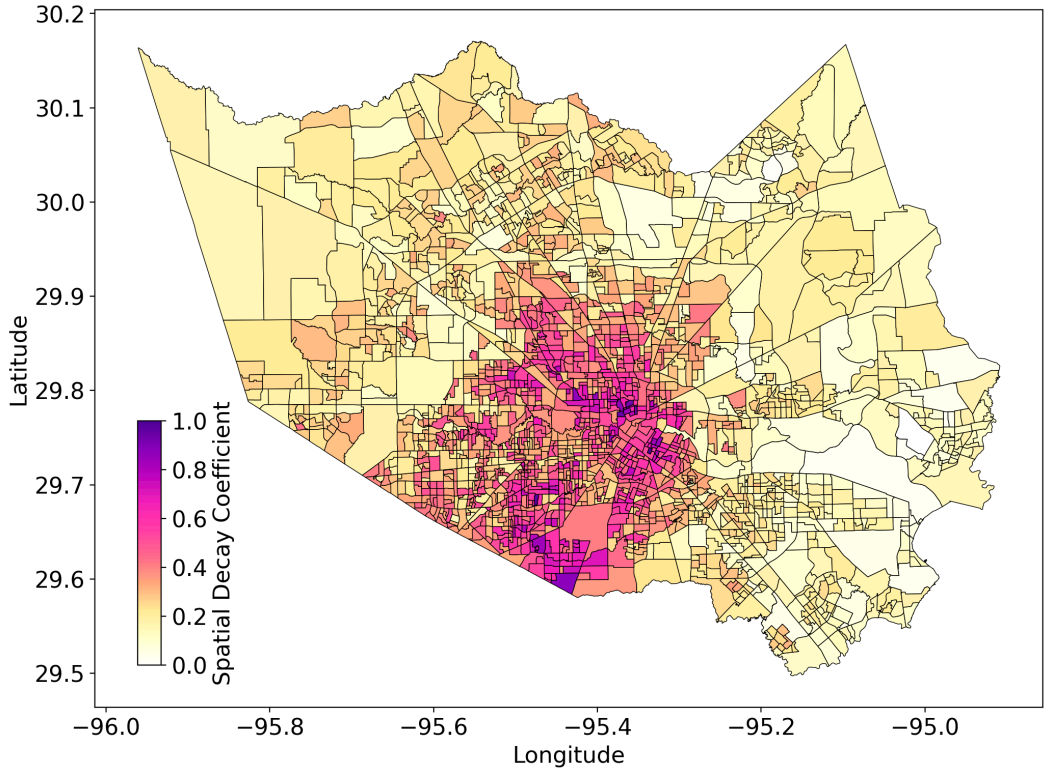

Figure 6. Spatial distribution of spatial decay coefficients obtained from the spatial decay model.

To further explore the spatial heterogeneity of spatial decay coefficients at the CBG level, we conducted a heterogeneity test. This test utilizes boxplots and compares cumulative distribution function (CDF) to examine the relationship between spatial decay coefficients and key urban structure features, such as POI density and road density, which are also control variables in our study. Figure 7a compares the CDF of spatial decay coefficients for CBGs grouped by low and high POI density. The trends across distributions are basically consistent, showing that the majority of the CBGs (more than 80%) exhibit low spatial decay coefficients (less than 0.5). Subsequently, the slope of the distributions starts to increase exponentially, indicating a drastically greater spatial decay coefficient for a smaller proportion of CBGs (less than 20%). This long-tail distribution of spatial decay coefficients underscores the disparity in community recovery within each group. When comparing the two distributions, the CDF for the low POI density group consistently sits above that of the high POI density group, suggesting that CBGs with lower POI density tend to have higher spatial decay coefficients. A one-way ANOVA test conducted in Figure 7b to prove the inter-group differences confirms a statistically significant distinction between the two POI density groups ($p < 0.001$***). The low POI density group has a higher spatial decay coefficient than that of the high POI density group. This result reveals not only a disproportional distribution of spatial decay coefficients among the POI density groups but also that the recovery speed of CBGs with low POI density has greater sensitivity to residential damage of surrounding CBGs compared with CBGs with high POI density.

We also conducted similar analyses for road density values versus the spatial decay coefficients, as presented in Figure 7c. The result shows that more CBGs with a high spatial decay coefficient can be identified from low road density groups than those from high road density groups. The ANOVA-test result in Figure 7d shows a significant level of group difference ($p < 0.001$***) in terms of spatial decay coefficient between the two road density groups. The low-road-density group is more at risk of experiencing community slower recovery speed due to the spillover effects of residential flood damage in surrounding areas than CBGs within a high-road-density group. These results reveal the heterogeneity of spatial spillover effects of residential damage on community recovery speed. The variations in the spatial decay coefficients can be explained based on the features of urban structure related to POI and road density. In other words, areas with a greater density of POIs and road density can represent city hubs with a greater concentration of activities and attractions. The recovery speed of these areas has less sensitivity to spillover effects of residential flood damage in farther neighborhoods. These findings shed light on the role of features related to urban structure in affecting the spatial diffusion of spillover effects of residential flood damage on recovery speed across different areas.

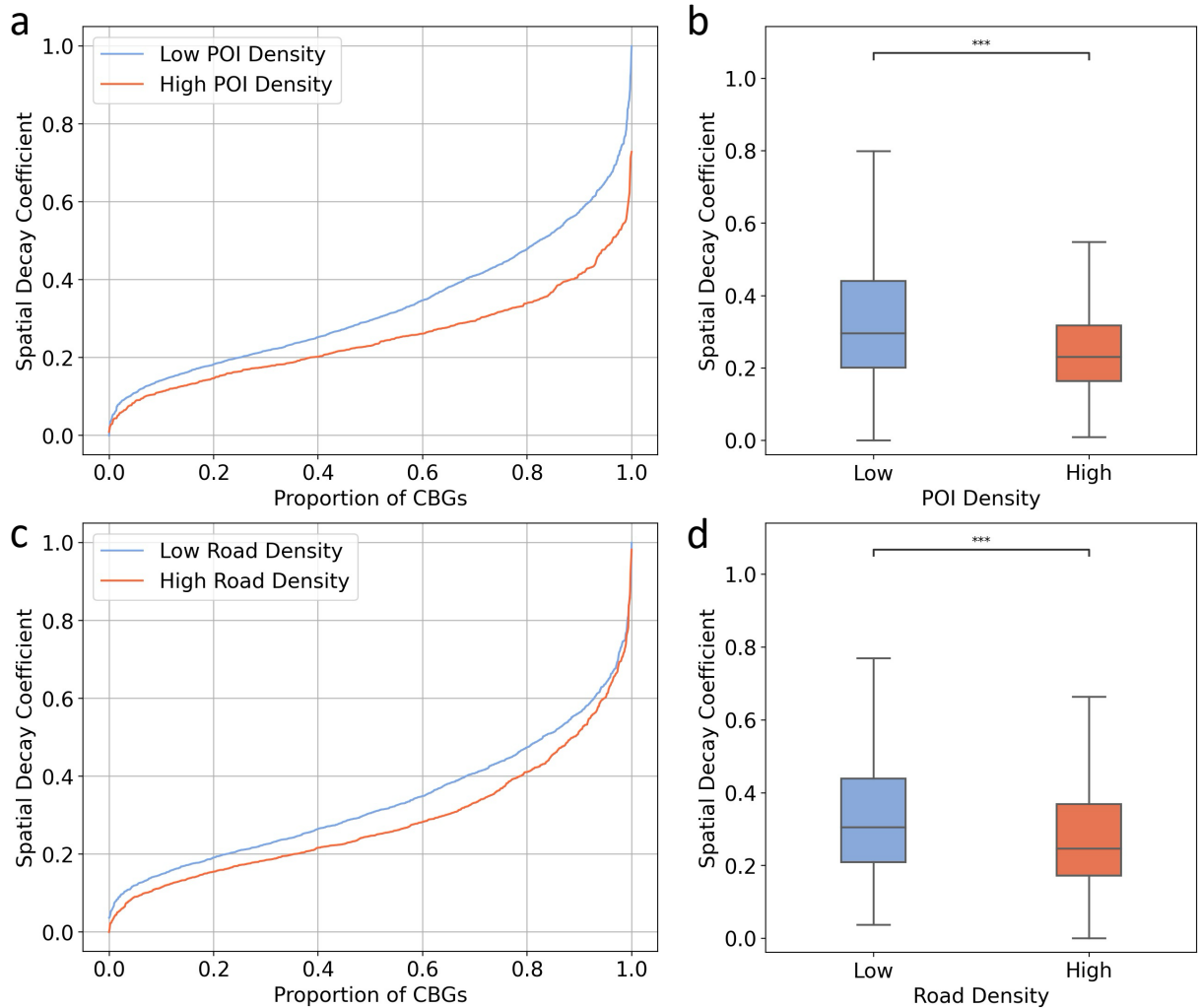

Figure 7. Distribution of spatial decay coefficients among groups of urban structure features. (a) CDF plot for low-POI-density and high-POI-density groups. (b) Boxplot showing the difference of spatial decay coefficients between low-POI-density and high-POI-density groups. (c) CDF plot for low-road-density and high road density groups. (d) Boxplot showing the difference of spatial decay coefficient between low-road-density and highroad-density groups. Note: *** $p <0.001$. We classified the POI density and road density into two groups based on the mean values. Statistically significant difference between groups was tested using one-way ANOVA.

# 5. Discussion and concluding remarks

This study investigated the spatial spillover effect, along with spatial reach and spatial decay of vulnerability of the built environment (i.e., residential flood damage) on community recovery after disasters. Departing from the existing literature that lacks empirical evidence regarding the characteristics of spatial diffusion processes that extend the effects of residential flood damage on community recovery speed beyond local impacts, this study used high-resolution datasets in spatial analyses to examine the extent to which the proximity to and severity of residential flood damage correlate with the recovery rates across different areas of an affected community.

One of the key novel aspects of this study is the use of unique high-resolution flood damage claim data to accurately determine and quantify residential flood damage extent. The vulnerability of the built environment is primarily evident through residential property damage during disasters [37]. Typically, property damage is quantified in terms of physical loss [38], and while many studies rely on the physical structure features (e.g., water depth) retrieved from the National Flood Insurance Program [21-23], these measures alone are insufficient for reliable determination of damage extent. Our research innovates by integrating NFIP and Individual Assistance claims to capture both insured and uninsured losses for broader coverage and by calculating property damage extent relative to the current market values of the properties. In addition, the use of detailed human mobility data enabled a precise quantification of the community recovery rate by employing population activities as a proxy for community functionality. We utilized these unique high-resolution datasets to develop a local regression model (ordinary least squares model) and three spatial models (spatial spillover model, spatial reach model, and spatial decay model) in Harris County, Texas, in the context of the 2017 Hurricane Harvey. The integration of these two high-resolution datasets enabled examination of the characteristics of the spatial diffusion process related to residential flood damage effects on community recovery.

The results highlight important characteristics related to the spatial diffusion of residential flood damage impacts on community recovery. By employing OLS and SLX models, we determined the impact of both the local and surrounding areas' residential flood damage on community recovery patterns. We found that local indicators of PDE (i.e., NC, MP, SDP, and MDP) demonstrate negative correlations with community recovery rate, indicating that more extensive and widespread damage correlates with slower recovery speed. Our findings also show that the number of claims is a more significant feature of built-environment vulnerability in terms of its spatial spillover effects on the recovery of other areas. The findings confirm the non-locality and spillover effects of residential flood damage on the recovery speed of areas farther away from the damaged areas. These insights underscore the importance of considering the spillover effects of residential property damage extent in developing regional recovery strategies, rather than focusing solely on the areas with the most severe damage.

Another important outcome of this study is the quantification of the spatial reach of residential flood damage on recovery rates of other areas. By setting varying spatial distance thresholds, we determined the specific reach of spatial spillover effects, addressing a common oversight in previous studies [39-41]. The results show that the influence of residential damage extends beyond the immediate area, affecting the recovery speed of nearby regions up to 31.2 miles away. In addition, while the impact of residential damage on recovery speed decreases with distance from the most affected areas, the uniformity of this impact varies. We proposed a novel spatial decay model inspired by the economic discounting decay model to measure the decay rate of spatial spillover effects. The model's results show that CBGs with high spatial decay coefficients are predominantly located in downtown Houston and its southwestern areas, suggesting that recovery speed in these areas is less sensitive to residential property damage in farther areas. The recovery rates of areas with higher concentrations of POIs and more developed road networks tend to have less sensitivity to residential flood damage in farther areas (and more sensitive to the damage of immediate surrounding areas). These findings reveal the heterogeneous nature of the spillover effects and the role of urban structure features in explaining this variation in the sensitivity of recovery rates in an area to the residential flood damage of farther areas. These findings bridge the gap between features of urban structure and spatial processes in community resilience [42]. The consideration of the role of urban structures in characterizing spatial diffusion processes that shape community resilience is essential for integrated urban design strategies that embed resilience characteristics through urban development plans.

In sum, the findings obtained in this study have multiple scientific contributions and practical implications. First, the findings reveal crucial aspects of how the built environment's vulnerability influences community recovery through spatial diffusion. This insight enhances our understanding of the extent to which vulnerabilities in the built environment can extend their impact from local areas to regional scales, influencing recovery speeds through spatial diffusion. In addition, the study identifies varied spillover effects and their sensitivity to urban structure features, offering insights into how urban planning can amplify or alleviate these effects to influence community resilience. This insight can inform integrated urban design strategies that could enhance regional resilience through urban development planning, such as facility distribution [43]. Collectively, the novel insights inform interdisciplinary fields of civil engineering, urban science, geography, and disaster science regarding spatial diffusion processes that shape community resilience. Second, the study utilizes unique, high-resolution datasets to specify the extent of residential damage and to assess community recovery speeds. The analysis incorporates both NFIP and IA flood claims to capture a broader scope of residential damage, including uninsured losses, which are often overlooked but were significant during events like Hurricane Harvey. The use of detailed human mobility data also allows for precise quantification of community recovery, employing population activities as a proxy for community functionality. The outcomes show the value of emerging high-resolution datasets in better characterization of the dynamics of community resilience. Third, the findings equip emergency managers and public officials with valuable insights into the non-local impacts of residential damage. This insight enables a more informed evaluation of flood risk reduction measures and their benefits for community resilience, taking into account the spatial spillover effects of local built-environment vulnerability on regional community resilience patterns.

**Declarations of interest**

The authors declare no competing interests.

**Acknowledgement**

This material is based in part upon work supported by the National Science Foundation under CRISP 2.0 Type 2, No. 1832662 grant and Texas A&M University X-Grant 699. The authors would like to acknowledge the data support from Spectus, Inc. and SafeGraph, Inc. The authors also thank the undergraduate researcher Vasudev Agarwal (Texas A&M University) for assistance with data processing. Any opinions, findings, conclusions, or recommendations expressed in this material are those of the authors and do not necessarily reflect the views of National Science Foundation, Texas A&M University, Spectus, Inc., or SafeGraph, Inc.

**Data availability**

Spectus data and SafeGraph data were collected through a CCPA- and GDPR-compliant framework and utilized for research purposes. The data that support the findings of this study are available from Spectus, Inc. and SafeGraph, Inc., but restrictions apply to the availability of these data, which were used under license for the current study. The data can be accessed upon request submitted on spectus.ai and safegraph.com. The flood damage claims data were provided by Dr. Samuel Brody and Dr. Russell Blessing (Texas A&M University at Galveston). Other data we used in this study are all publicly available.

**Code availability**

The code that supports the findings of this study is available from the corresponding author upon request.

**Author contributions**

J.M.: Conceptualization, Methodology, Data curation, Formal analysis, Writing—Original draft. A.M.: Conceptualization, Methodology, Writing—Reviewing and Editing, Supervision, Funding acquisition. R.B.: Data curation, Writing—Reviewing and Editing. S.B.: Data curation, Writing—Reviewing and Editing.

**Additional information**

Supplementary information associated with this article can be found in the Supplementary Information document.

## References

1. Yabe, T., et al., *Toward data-driven, dynamical complex systems approaches to disaster resilience.* Proceedings of the National Academy of Sciences, 2022. **119**(8): p. e2111997119.
2. Yang, Y., et al., *A federated pre-event community resilience approach for assessing physical and social sub-systems: An extreme rainfall case in Hong Kong.* Sustainable Cities and Society, 2020. **52**: p. 101859.
3. McPhearson, T., et al., *Advancing understanding of the complex nature of urban systems*. 2016, Elsevier. p. 566-573.
4. Li, L., et al., *Post-disaster functional recovery of the built environment: A systematic review and directions for future research.* International Journal of Disaster Risk Reduction, 2023: p. 103899.
5. Nofal, O.M. and J.W. Van De Lindt, *Understanding flood risk in the context of community resilience modeling for the built environment: Research needs and trends.* Sustainable and Resilient Infrastructure, 2022. **7**(3): p. 171-187.
6. Golla, A.P.S., S.P. Bhattacharya, and S. Gupta, *Assessing the discrete and systemic response of the Built Environment to an earthquake.* Sustainable Cities and Society, 2022. **76**: p. 103406.
7. Liu, C.-F. and A. Mostafavi, *Network diffusion model reveals recovery multipliers and heterogeneous spatial effects in post-disaster community recovery.* Scientific Reports, 2023. **13**(1): p. 19032.
8. Ma, J., B. Li, and A. Mostafavi, *Characterizing urban lifestyle signatures using motif properties in network of places.* Environment and Planning B: Urban Analytics and City Science, 2024. **51**(4): p. 889-903.
9. Ma, J. and A. Mostafavi, *Decoding the Pulse of Community during Disasters: Resilience Analysis Based on Fluctuations in Latent Lifestyle Signatures within Human Visitation Networks.* arXiv preprint arXiv:2402.15434, 2024.
10. Rajput, A.A., et al., *Human-centric characterization of life activity flood exposure shifts focus from places to people.* Nature Cities, 2024: p. 1-11.
11. Fan, C., X. Jiang, and A. Mostafavi, *Evaluating crisis perturbations on urban mobility using adaptive reinforcement learning.* Sustainable Cities and Society, 2021. **75**: p. 103367.
12. Coleman, N., et al., *Lifestyle pattern analysis unveils recovery trajectories of communities impacted by disasters.* Humanities and Social Sciences Communications, 2023. **10**(1): p. 1-13.
13. Llaguno-Munitxa, M. and E. Bou-Zeid, *The environmental neighborhoods of cities and their spatial extent.* Environmental Research Letters, 2020. **15**(7): p. 074034.
14. Truelove, H.B., et al., *Positive and negative spillover of pro-environmental behavior: An integrative review and theoretical framework.* Global Environmental Change, 2014. **29**: p. 127-138.
15. Jiang, X., W. Fu, and G. Li, *Can the improvement of living environment stimulate urban innovation?——analysis of high-quality innovative talents and foreign direct investment spillover effect mechanism.* Journal of Cleaner Production, 2020. **255**: p. 120212.
16. Westlake, S., C.H. John, and E. Cox, *Perception spillover from fracking onto public perceptions of novel energy technologies.* Nature Energy, 2023. **8**(2): p. 149-158.
17. Paroussos, L., et al., *Climate clubs and the macro-economic benefits of international cooperation on climate policy.* Nature Climate Change, 2019. **9**(7): p. 542-546.
18. Fan, C., T. Yabe, and T. Liu, *Quantifying Complex Urban Spillover Effects via Physics-based Deep Learning.* 2023.
19. Service, N.W., *Hurricane Harvey & Its Impacts on Southeast Texas (August 25-29, 2017).* https://www.weather.gov/hgx/hurricaneharvey, 2017.
20. Furrh, J.T. and P. Bedient, *Upstream Addicks–Barker reservoir damages during Hurricane Harvey: A case study of urban hydrology and policy failure in Houston, TX.* JAWRA Journal of the American Water Resources Association, 2023. **59**(5): p. 984-998.

Supplementary Information for

*Non-locality and Spillover Effects of Residential Flood Damage on Community Recovery: Insights from High-resolution Flood Claim and Mobility Data*

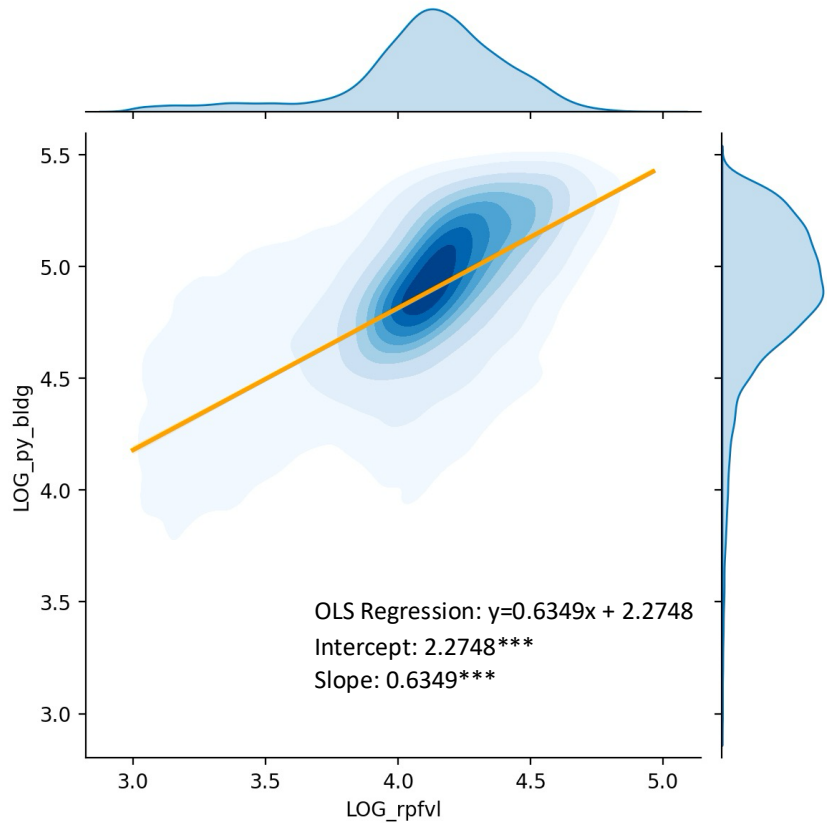

Figure S1. Density distribution and quantitative linear relationship between the NFIP and IA datasets using an OLS linear regression method. Both data were logged, with LOG_rpfvl representing real property damage assessments by FEMA in the IA and LOG_py_bldg representing building claim payments from NFIP. By this method, we normalized the NFIP and IA data to the same scale.

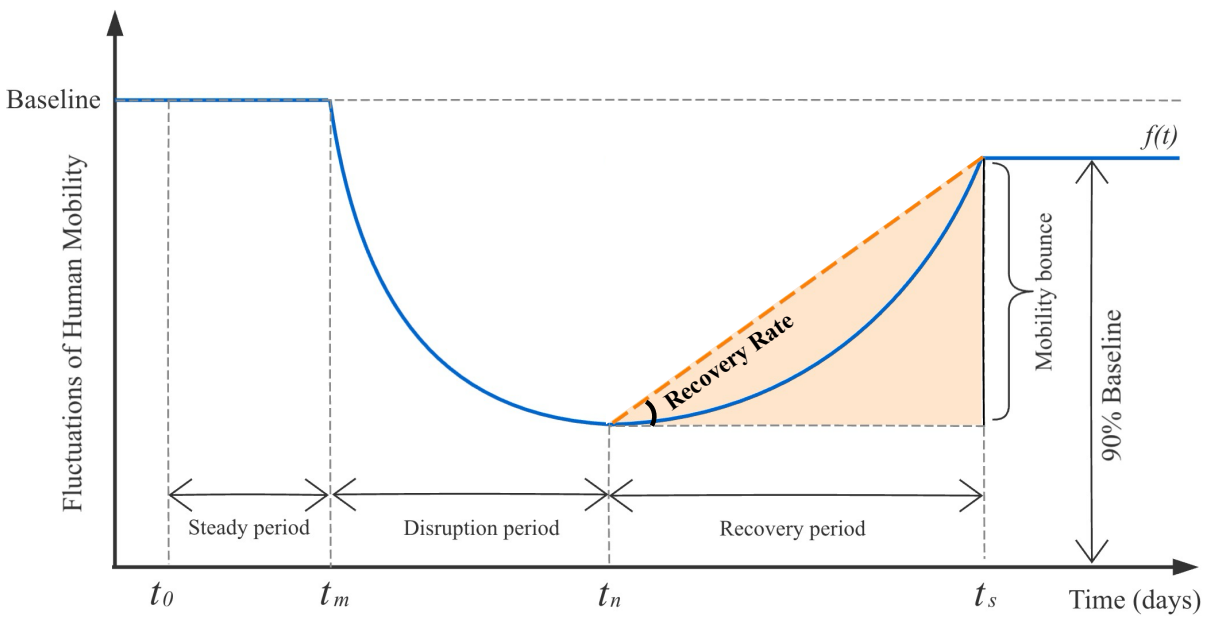

Figure S2. Conceptual framework for describing the trend of percent change of movement rate over time during a hurricane.

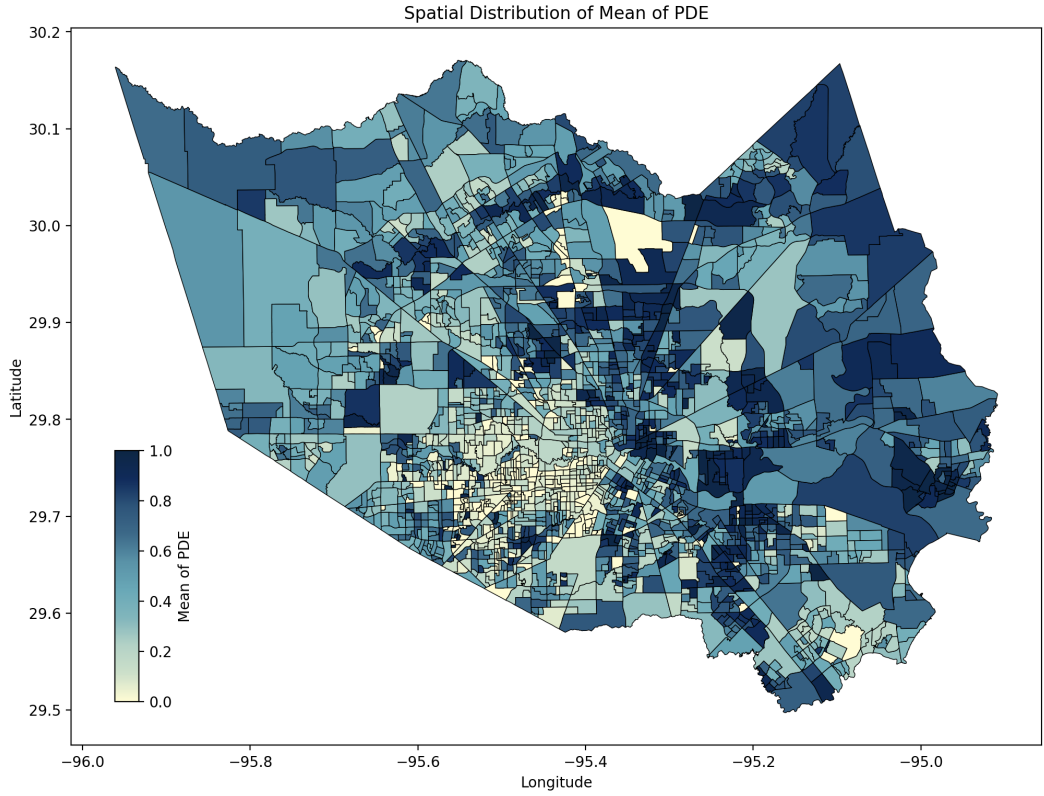

Figure S3. Spatial distribution of mean of PDE at the CBG level.

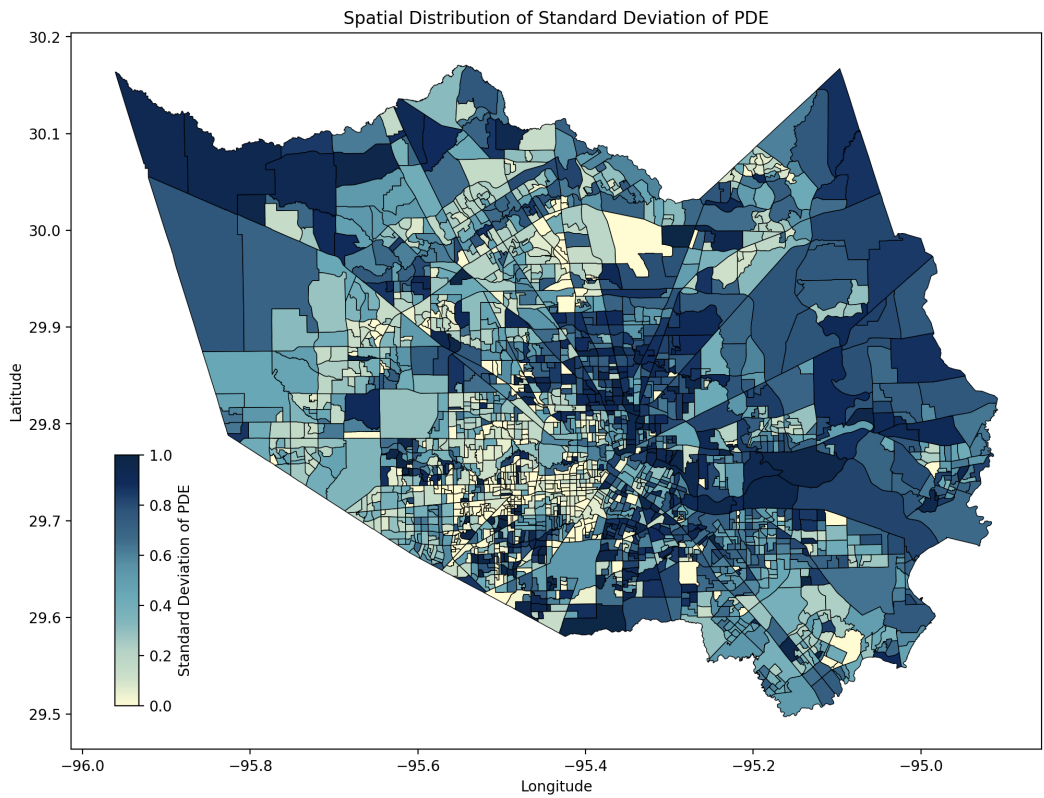

Figure S4. Spatial distribution of standard deviation of PDE at the CBG level.

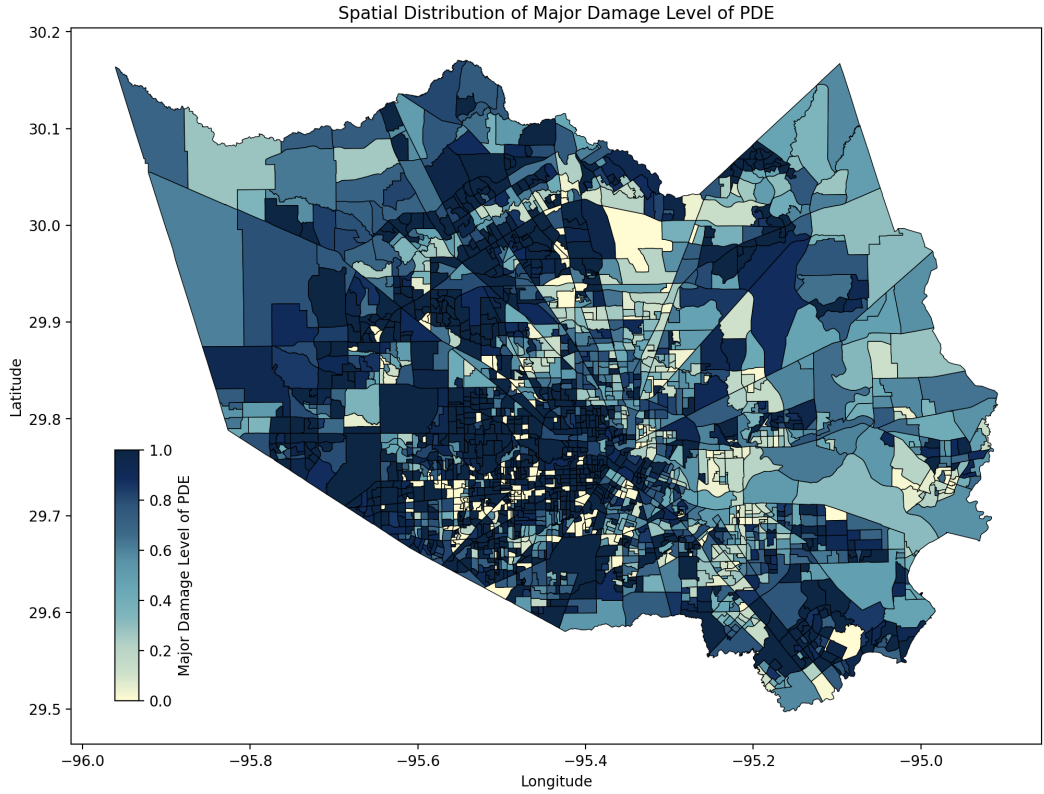

Figure S5. Spatial distribution of major damage level of PDE at the CBG level.

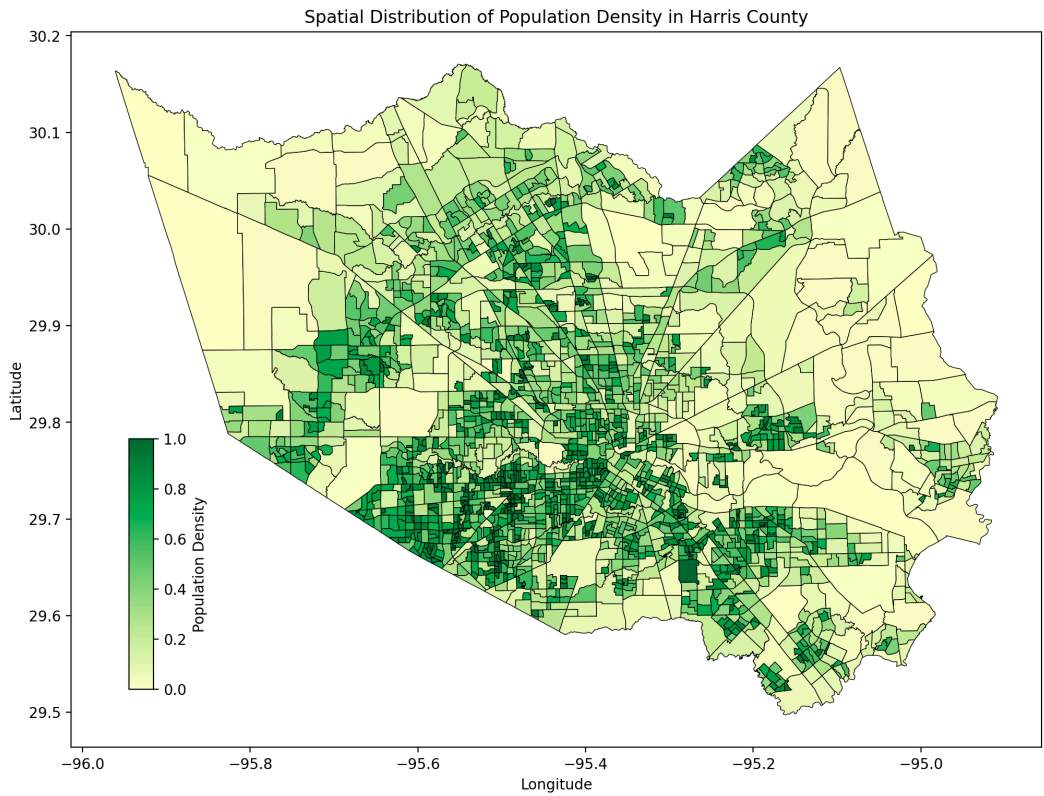

Figure S6. Spatial distribution of population density at the CBG level.

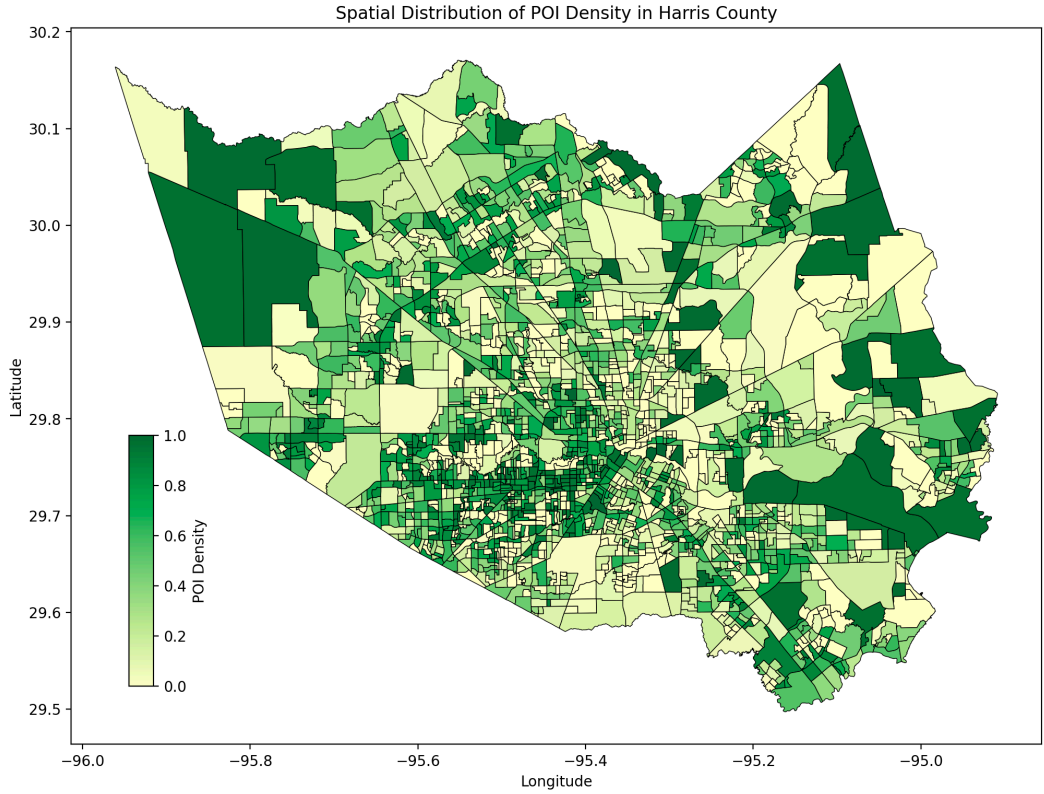

Figure S7. Spatial distribution of POI density at the CBG level.

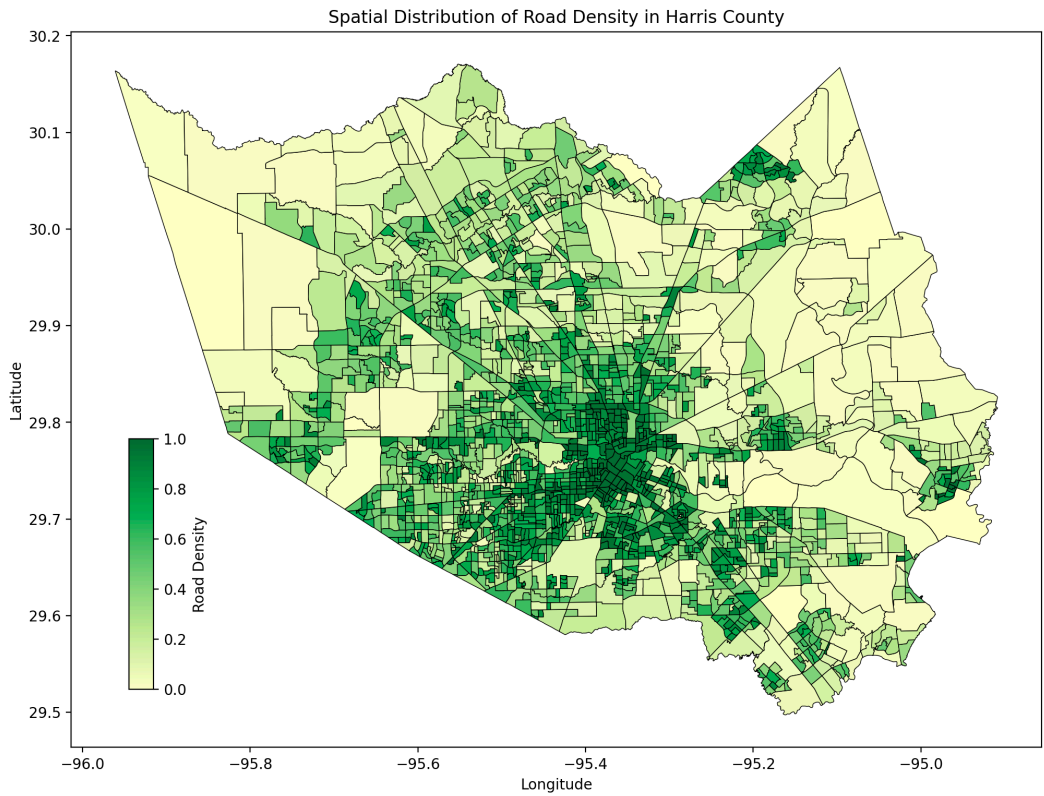

Figure S8. Spatial distribution of road density at the CBG level.

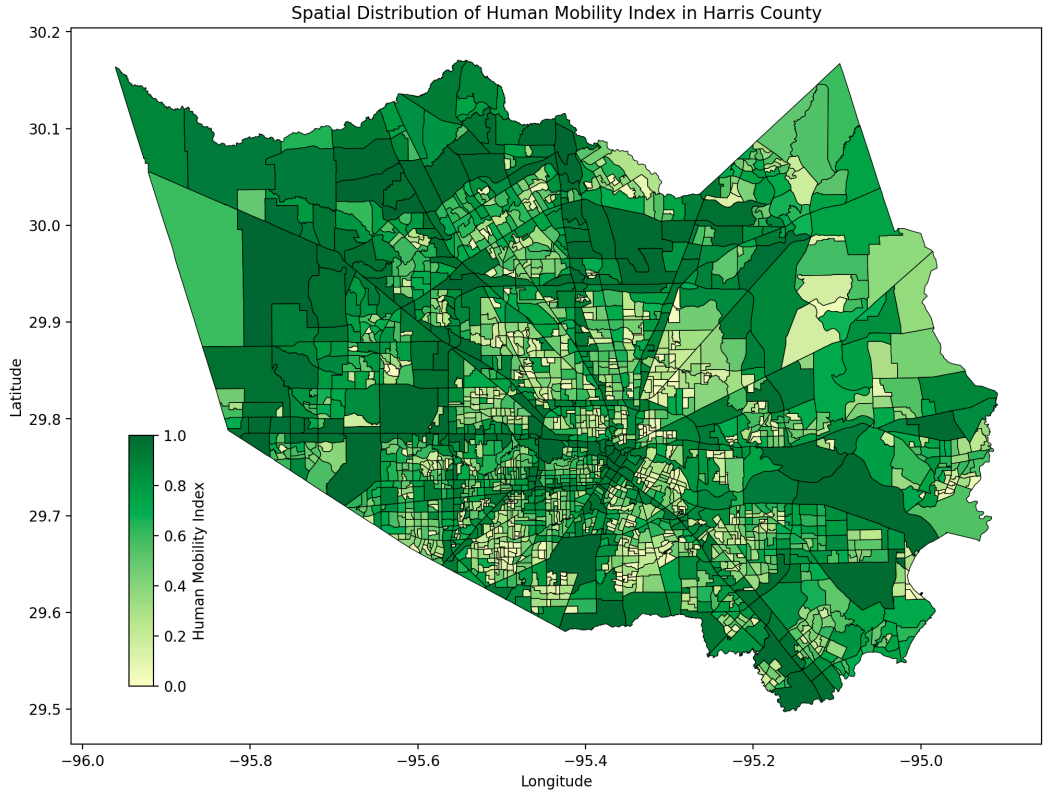

Figure S9. Spatial distribution of human mobility index at the CBG level.

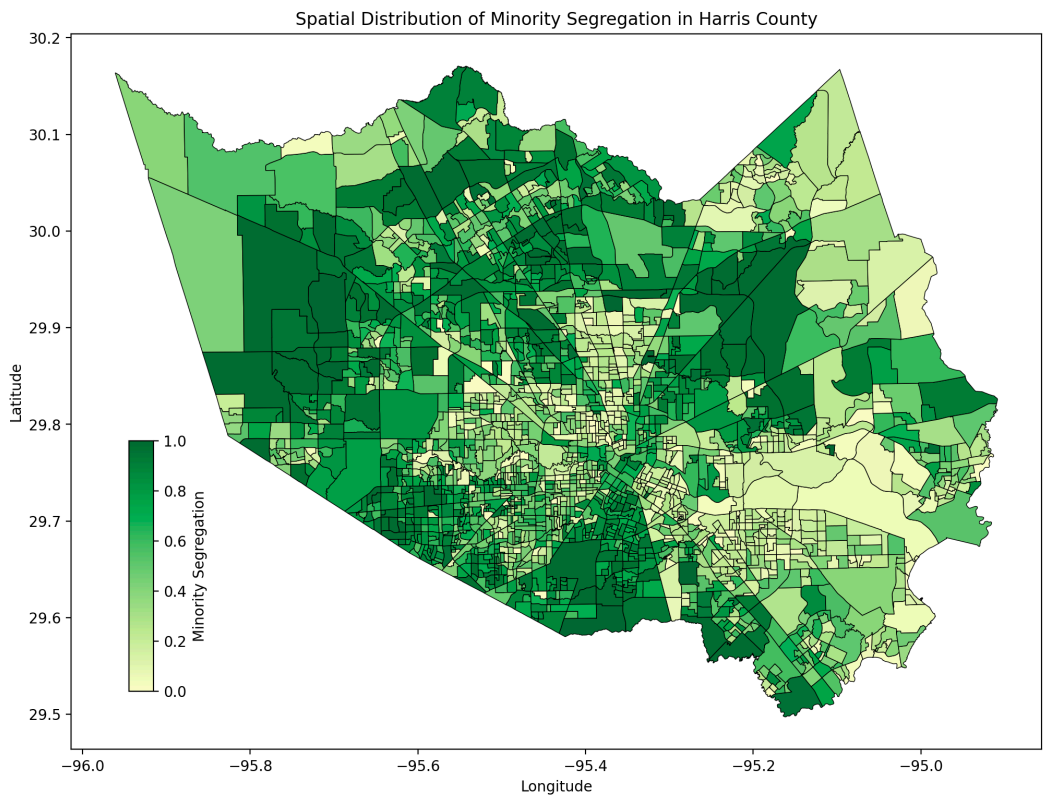

Figure S10. Spatial distribution of minority segregation at the CBG level.

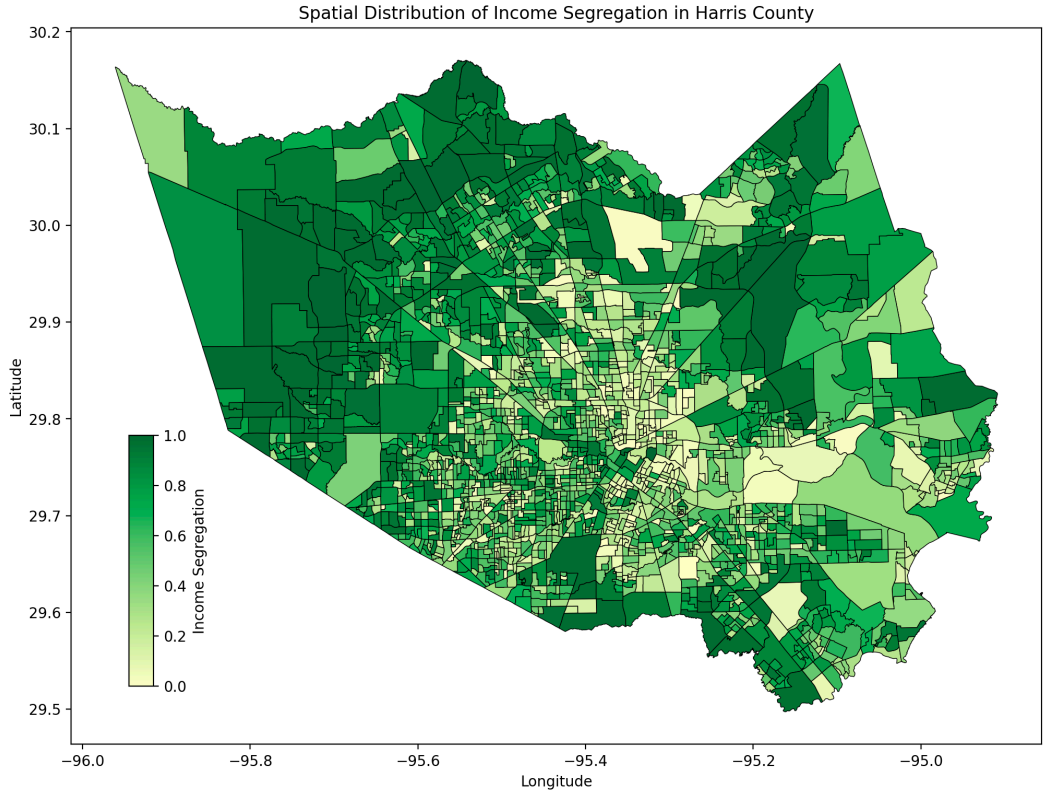

Figure S11. Spatial distribution of income segregation at the CBG level.

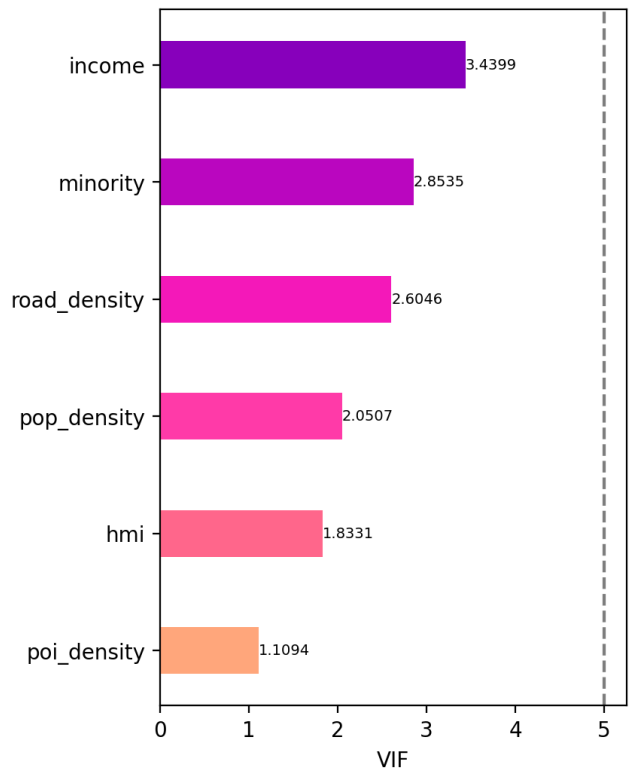

Figure S12. Multicollinearity test results of the control variables using Variance Inflation factor (VIF).

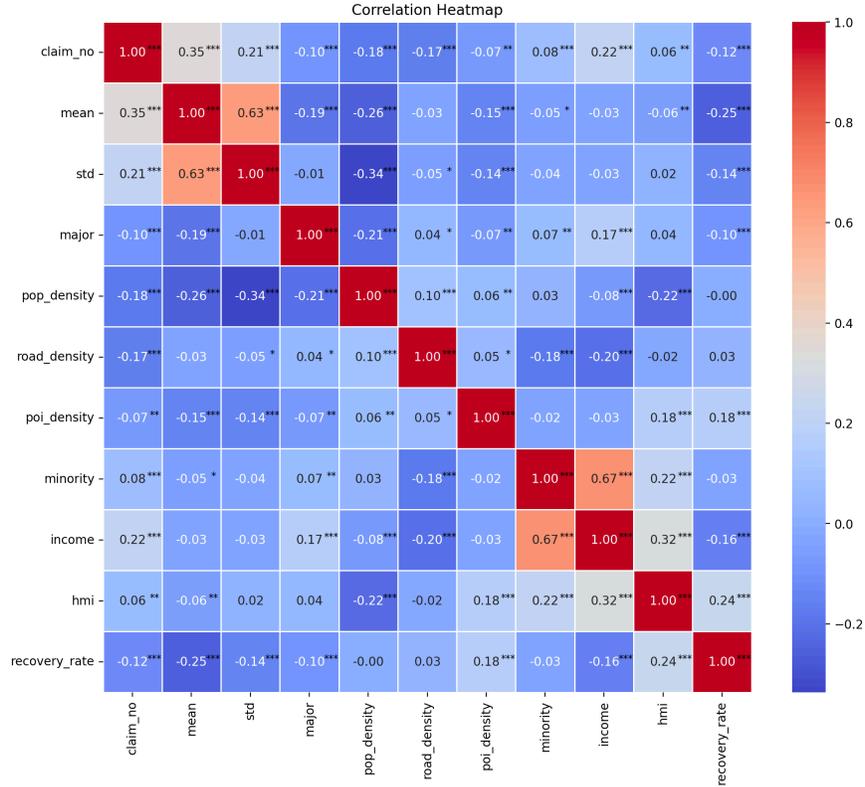

Figure S13. Pearson correlation heatmap for the property damage extent features and control variables.

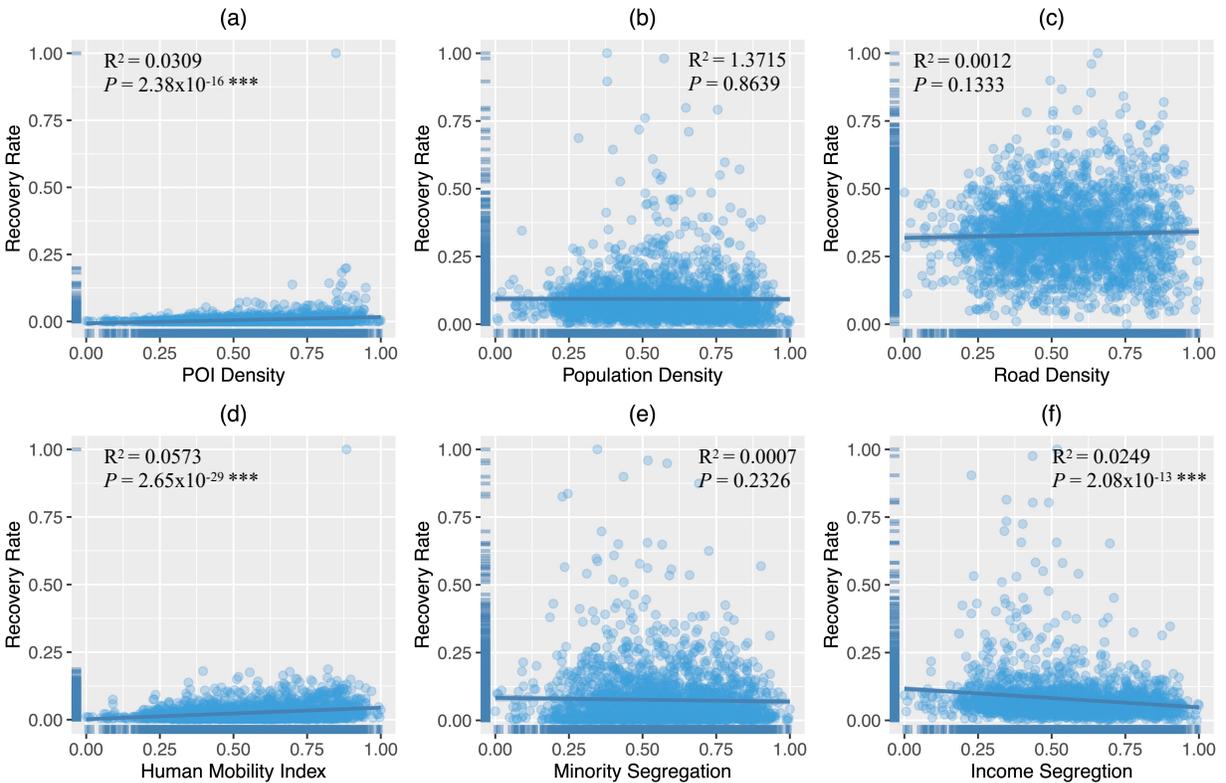

Figure S14. Correlation between control variables and recovery rate.

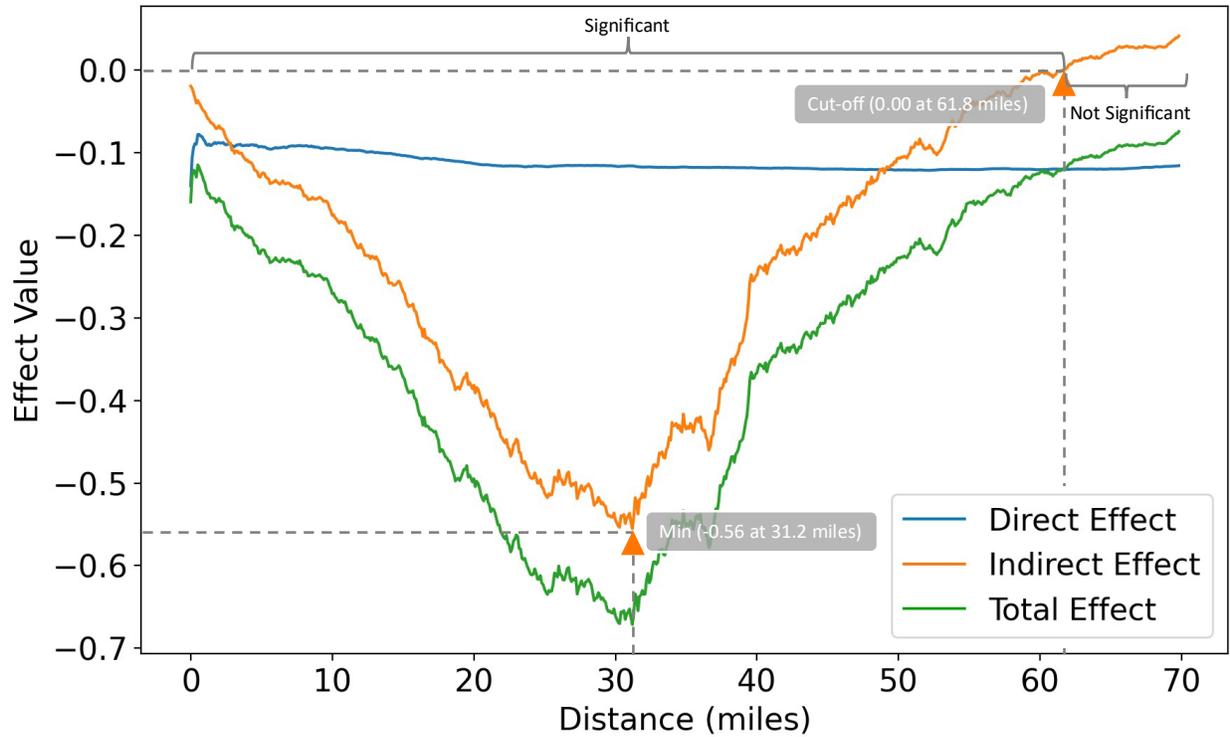

Figure S15. Decomposition effects of the mean of PDE as the distance threshold changes in the spatial weight matrix.

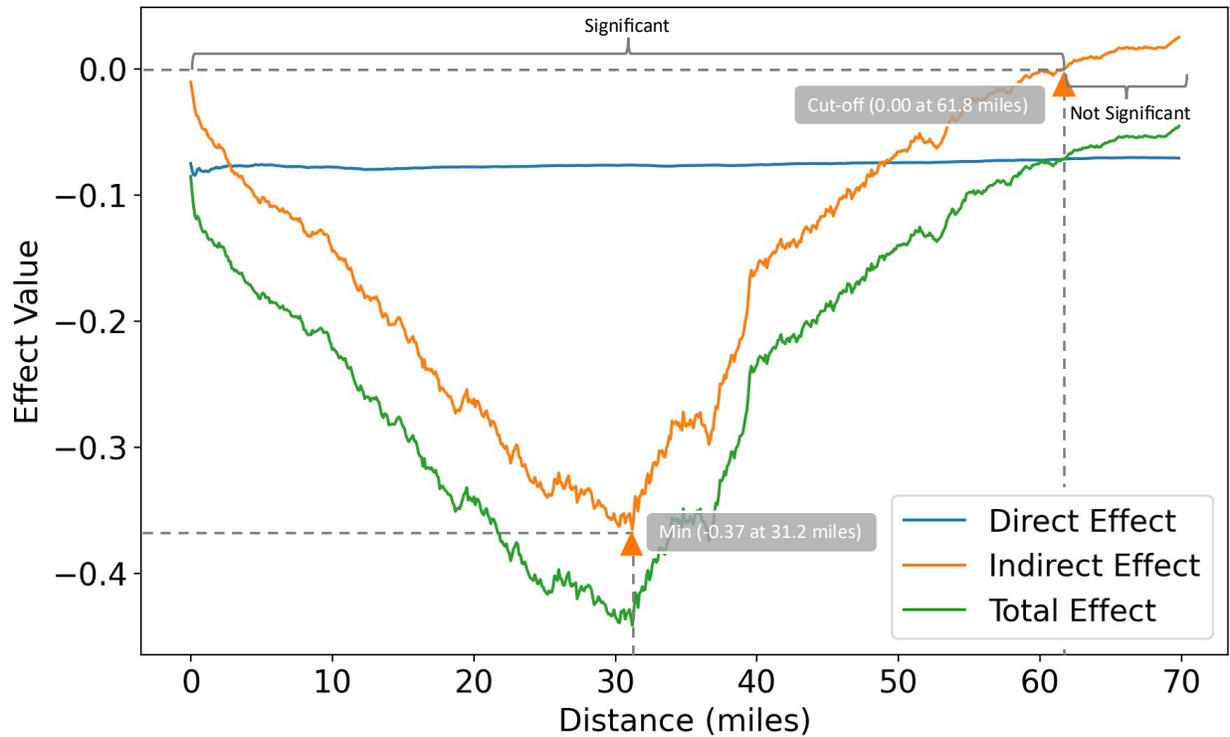

Figure S16. Decomposition effects of the major damage level of PDE as the distance threshold changes in the spatial weight matrix.

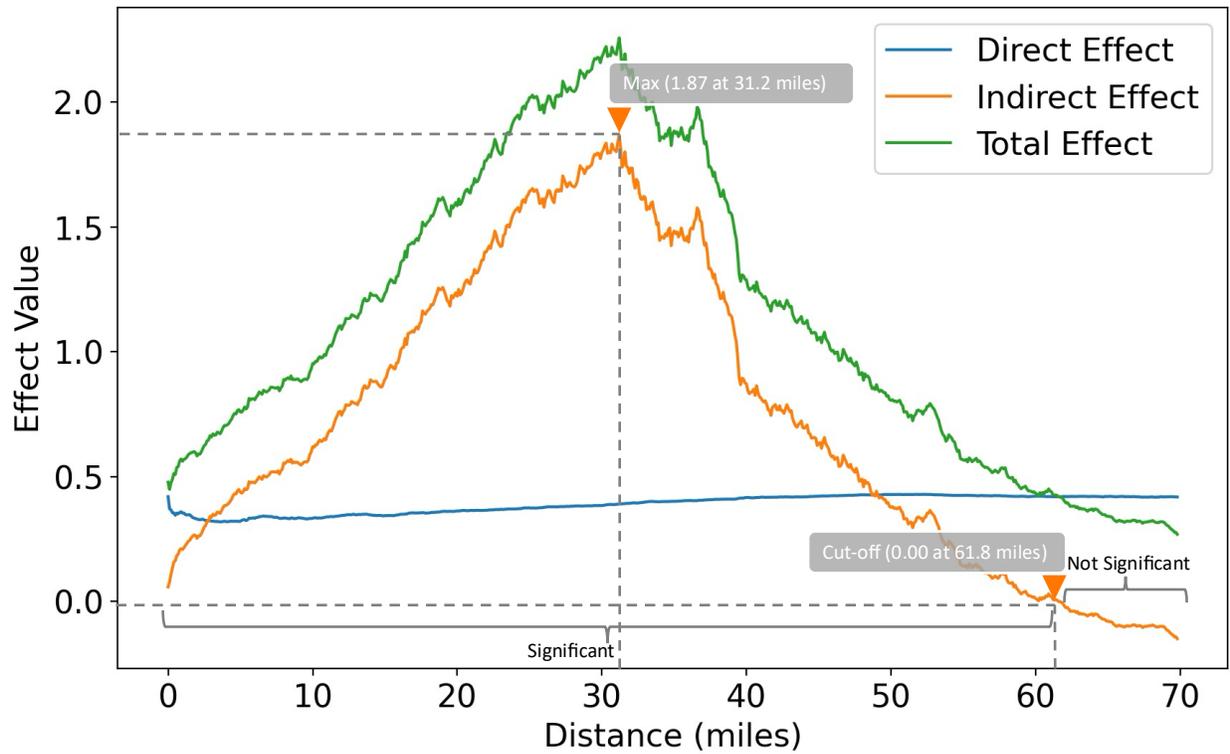

Figure S17. Decomposition effects of the human mobility index as the distance threshold changes in the spatial weight matrix.

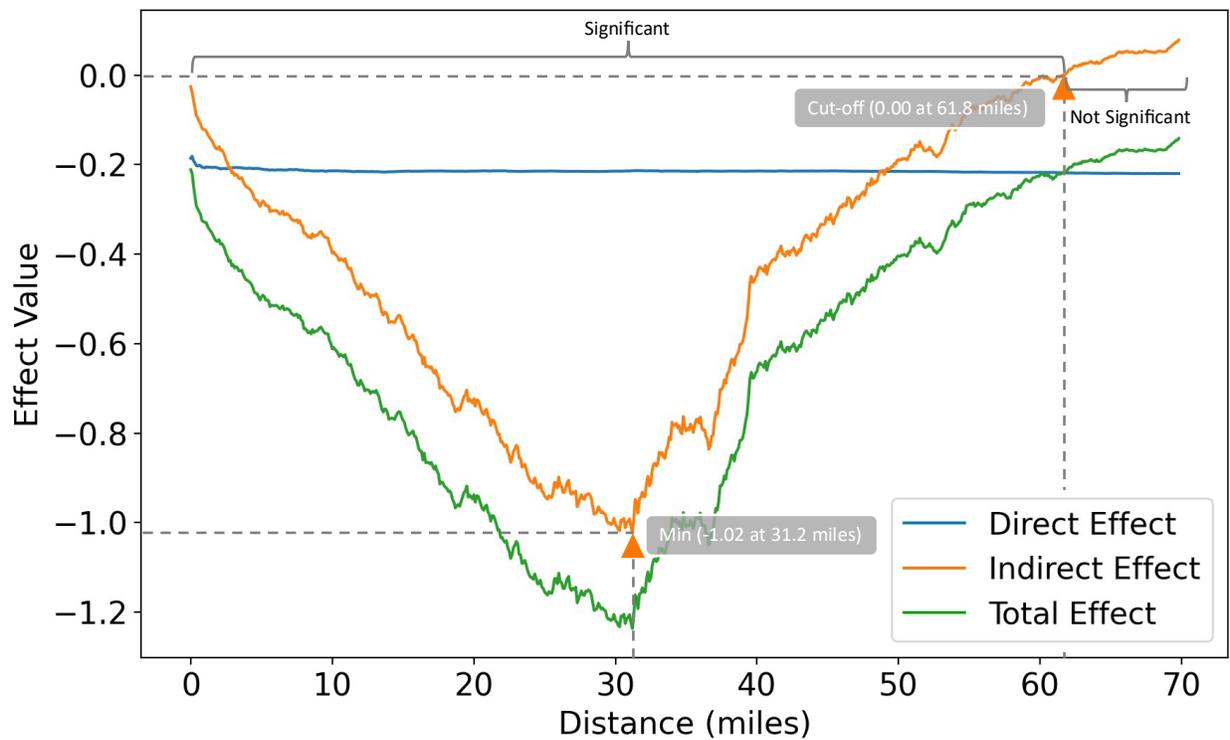

Figure S18. Decomposition effects of the road density as the distance threshold changes in the spatial weight matrix.

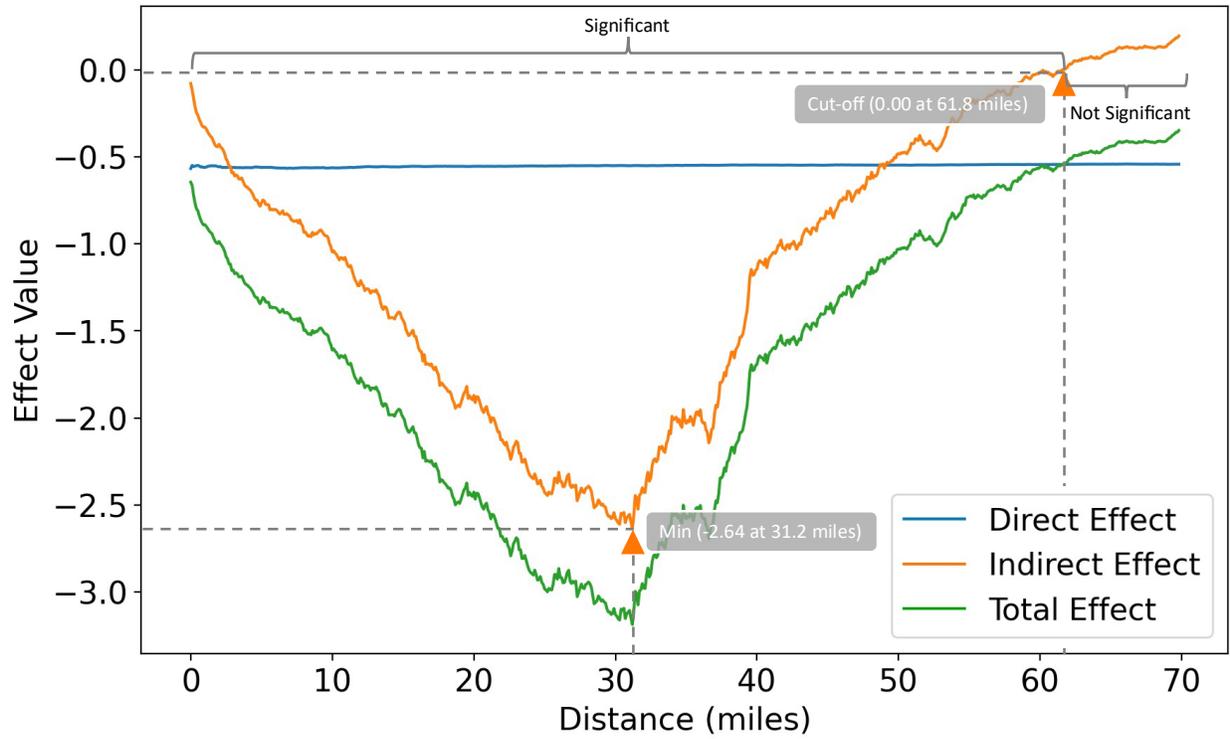

Figure S19. Decomposition effects of the income segregation as the distance threshold changes in the spatial weight matrix.

Table S1. Regression result and model performance of SLX models with different spatial weight matrixes.

| Variables | Shared Boundary (Global Moran's I: 0.127, P: 0.001***) | | | | K Nearest Neighbors (k=5) (Global Moran's I: 0.093, P: 0.001***) | | | | Inverse Square of Geographical Distance (Global Moran's I: 0.143, P: 0.001***) | | | |
|---|---|---|---|---|---|---|---|---|---|---|---|---|
| | Coefficient | Direct | Indirect | Total | Coefficient | Direct | Indirect | Total | Coefficient | Direct | Indirect | Total |
| NC | -0.218** | -0.112*** | -0.192*** | -0.304*** | -0.321*** | -0.212*** | -0.233** | -0.445*** | -0.333** | -0.229*** | -0.290*** | -0.519*** |
| MP | -0.277* | -0.128** | -0.176** | -0.304** | -0.257** | -0.268*** | -0.238** | -0.506** | -0.377* | -0.210*** | -0.238** | -0.448** |
| SDP | 0.000 | 0.000 | -0.009 | -0.009 | -0.001 | 0.000 | 0.000 | 0.000 | -0.001 | 0.000 | -0.001 | -0.001 |
| MDP | -0.198*** | -0.002* | -0.006** | -0.008* | -0.100*** | -0.120* | -0.125** | -0.245* | -0.188** | -0.120* | -0.115** | -0.235* |
| POP | -0.008 | -0.001 | -0.003 | -0.004 | -0.124 | 0.000 | 0.002 | 0.002 | -0.078 | -0.120 | -0.159 | -0.279 |
| RD | -0.104** | -0.213** | -0.339** | -0.552** | -0.090* | -0.103** | -0.217*** | -0.320** | -0.099*** | -0.100** | -0.201*** | -0.301* |
| POI | 0.133* | 0.300** | 0.220*** | 0.520*** | 0.204*** | 0.156** | 0.198** | 0.354*** | 0.190* | 0.432*** | 0.430*** | 0.862** |
| MS | -0.101 | -0.006 | -0.035 | -0.041 | -0.001 | 0.000 | -0.011 | -0.011 | 0.000 | -0.016 | -0.015 | -0.031 |
| IS | -0.082** | -0.125*** | -0.171** | -0.296*** | -0.009*** | -0.005*** | -0.070* | -0.075*** | -0.299*** | -0.005*** | -0.150** | -0.155*** |
| HMI | 0.110*** | 0.009* | 1.305** | 1.314*** | 0.110*** | 0.078* | 0.105** | 0.183*** | 0.110*** | 0.008* | 0.005** | 0.013** |
| $R^2$ | 0.218 | | | | 0.312 | | | | 0.422 | | | |
| Adjusted $R^2$ | 0.208 | | | | 0.243 | | | | 0.319 | | | |
| Log-likelihood | 767.234 | | | | 777.098 | | | | 801.364 | | | |
| AIC | -1903.219 | | | | -1837.200 | | | | -1709.784 | | | |
| Cut-off distance | 56.0 (-9.39%) | | | | 66.5 (7.61%) | | | | 60.1 (-2.75%) | | | |
| Min/max distance | 27.3 (-12.5%) | | | | 36.8 (17.95%) | | | | 30.1 (-3.53%) | | | |

Note: Number of observation: 2, 144 CBGs; ***, **, and * refer to the significance level at 1%, 5%, and 10%, respectively. The values in the parentheses next to the cut-off distance and minimum/maximum distance values indicate deviation rates, which represent the deviation of these values from the corresponding cut-off distance and minimum/maximum distance values in the main text.

**Note S1. Definition and data for control variables**
**Population density:** The population size was obtained from the 2017 race and ethnicity data from US Census Bureau [1]. We calculated the population density at the CBG level by dividing the total population of the CBG by its land area. Land area data was also obtained from US Census Bureau [2].

**Minority segregation and income segregation:** Urban segregation refers to the physical and social separation of different racial, ethnic, and socioeconomic groups within a city [3]. This separation can take many forms, including minority segregation and income segregation. One of the key consequences of urban segregation is that it often leads to unequal distribution of resources, as well as increased exposure to environmental hazards such as flooding [4, 5].

In this study, we adopted the Dissimilarity Index (DI) to evaluate minority segregation and income segregation. The DI is a measure of spatial segregation that indicates the extent to which two groups are evenly distributed across different areas, which ranges from 0 (indicating perfect evenness) to 1 (indicating complete separation) [6, 7]. We calculated the DI based on the proportion of minority population (for minority segregation) and the proportion of low-income population (for income segregation) at the CBG level relative to the census tract level [8]:

$$DI = \frac{1}{2}\sum_{i=1}^{n}\left|\frac{x_i}{X} - \frac{y_i}{Y}\right| \quad (1)$$

where $x_i$ is the minority population (or low-income population) in the smaller geographical unit; $X$ is the the minority population (or low-income population) in the larger geographical unit; $y_i$ is the reference population in the smaller geographical unit; $Y$ is the reference population in the larger geographical unit. In our study, smaller geographical unit refers to CBG level and the larger geographical unit refers to census tract level.

For minority segregation, we collected the racial population data from the 2017 race and ethnicity data from US Census Bureau [1]. The primary racial groups in this study are non-Hispanic White, non-Hispanic Black, and non- Hispanic Asian residents. Non-Hispanic populations are selected because White, Black, or Asian populations can be mutually selective from Hispanic populations. The method of this kind of analysis is consistent with those frequently adopted by high-impact research works [9-11]. We considered the non-Hispanic Black, and non- Hispanic Asian as the minority population and non-Hispanic White as the reference population.

For income segregation, we extracted median income data from the 2017 American Community Survey (ACS) [12]. This study used the 5-year estimates of median income due to the broader coverage of areas, larger sample size, and higher precision, making the data more reliable than 1-year and 3-year estimates. We used the quantile income groups of a CBG (Q1 to Q4) to indicate income levels with Q1/Q2 representing low-income groups and Q3/Q4 represent high-income groups, respectively.

**POI density:** To capture the distribution of physical facilities, we adopted the POI data in the Harris County from SafeGraph [13]. The dataset includes basic information about POIs, such as POI IDs, location names, geographical coordinates, addresses, brands, and North American Industry Classification System (NAICS) codes to categorize POIs. The NAICS code is the standard used by federal statistical agencies in classifying business establishments [14]. In this study, we selected 10 essential types of POIs that are closely relevant to human daily lives: restaurants, schools, grocery stores, churches, gas stations, pharmacies and drug stores, banks, hospitals, parks, and shopping malls. We counted the number of POIs in each CBG and calculated its density as their facility distribution feature.

**Road density:** To capture the distribution of transportation network, we extracted data from Open Street Map [15] to calculate the density of road segments in CBGs. We estimated complete road networks from the raw data by assembling road segments. Since the lengths of road segments created by the source were in close proximity, we calculated road density by dividing the number of road segments by the area of a CBG.

**Human mobility index:** To understand the inequality of population activities, we employed mobile phone data from Spectus Inc. to develop the metric of human mobility index (HMI). The data has a wide set of attributes, including anonymized user ID, latitude, longitude, POI ID, time of observation, and the dwelling time of each visit [16]. Prior studies found that Spectus mobile phone data is representative to describe human activities and mobility [17-19]. Hence, the feature generated using the dataset should be representative and valid for our analyses. We extracted the data from April 2019 (28 days) to account for the variation of population activities on weekdays and weekends. Our period is also during regular conditions when no external extreme events perturbed human activities. To develop the HMI, we first assigned each visit point $v_i$ to a defined CBG in a county. Then, we calculated HMI as follows:

$$HMI = \frac{\sum_{i=1}^{n} v_i}{28} \tag{6}$$

where *n* denotes the number of visits in a CBG. We finally mapped the values of HMI to the range from 0 to 1 using min-max scaling. The proximity of HMI values to 0 or 1 indicates the level of human mobility and activity, with values closer to 0 indicating lower activity and values closer to 1 indicating higher activity in a CBG.

**References**

1. *US Census Bureau.* Hispanic or Latino, and not Hispanic or Latino by race. Census Bureau Data. , 2017. https://data.census.gov/cedsci/.
2. *US Census Bureau* USA Counties: 2011, 2011. https://www.census.gov/library/publications/2011/compendia/usa-counties-2011.html#LND.
3. Li, Q.-Q., et al., *Towards a new paradigm for segregation measurement in an age of big data.* Urban Informatics, 2022. 1(1): p. 5.
4. Martines, M.R., et al., *Spatial segregation in floodplain: An approach to correlate physical and human dimensions for urban planning.* Cities, 2020. 97: p. 102551.
5. Moro, E., et al., *Mobility patterns are associated with experienced income segregation in large US cities.* Nature communications, 2021. 12(1): p. 4633.
6. Massey, D.S. and N.A. Denton, *The dimensions of residential segregation.* Social forces, 1988. 67(2): p. 281-315.
7. Lichter, D.T., et al., *National estimates of racial segregation in rural and small-town America.* Demography, 2007. 44(3): p. 563-581.
8. Kodros, J.K., et al., *Unequal airborne exposure to toxic metals associated with race, ethnicity, and segregation in the USA.* Nature Communications, 2022. 13(1): p. 6329.
9. Liu, T. and C. Fan, *Impacts of disaster exposure on climate adaptation injustice across US cities.* Sustainable Cities and Society, 2023. 89: p. 104371.
10. Jbaily, A., et al., *Air pollution exposure disparities across US population and income groups.* Nature, 2022. 601(7892): p. 228-233.
11. Mehta, N.K., H. Lee, and K.R. Ylitalo, *Child health in the United States: recent trends in racial/ethnic disparities.* Social Science & Medicine, 2013. 95: p. 6-15.
12. *US Census Bureau.* Income in the past 12 months (in 2017 Inflation-Adjusted Dollars). 2017. https://data.census.gov/cedsci/.
13. *SafeGraph.* 2023. https://www.safegraph.com/.
14. *US Census Bureau* North American Industry Classification System, 2017. https://www.census.gov/naics/.
15. *Open Street Map.* 2023. https://www.openstreetmap.org.
16. *Spectus.* 2023. https://spectus.ai/.
17. Aleta, A., et al., *Modelling the impact of testing, contact tracing and household quarantine on second waves of COVID-19.* Nature Human Behaviour, 2020. 4(9): p. 964-971.
18. Fan, C., et al., *Interpretable machine learning learns complex interactions of urban features to understand socio-economic inequality.* Computer-Aided Civil and Infrastructure Engineering, 2023.
19. Wang, F., et al., *Extracting trips from multi-sourced data for mobility pattern analysis: An app-based data example.* Transportation Research Part C: Emerging Technologies, 2019. 105: p. 183-202.